# New infrared spectra of CO₂ – Xe: modeling Xe isotope effects, intermolecular bend and stretch, and symmetry breaking of the CO₂ bend

A.J. Barclay,[1] A.R.W. McKellar,[2] Colin M. Western,[3] and N. Moazzen-Ahmadi[1]

[1] *Department of Physics and Astronomy, University of Calgary, 2500 University Drive North West, Calgary, Alberta T2N 1N4, Canada*

[2]*National Research Council of Canada, Ottawa, Ontario K1A 0R6, Canada*

[3]*School of Chemistry, University of Bristol, Bristol BS8 1TS, U.K.*

**Abstract**

The infrared spectrum of the weakly-bound $CO_2$-Xe complex is studied in the region of the carbon dioxide $\nu_3$ fundamental vibration ($\approx$2350 cm$^{-1}$), using a tunable OPO laser source to probe a pulsed supersonic slit jet expansion. The Xe isotope dependence of the spectrum is modeled by scaling the vibrational and rotational parameters, with the help of previous microwave data. The scaling model provides a good simulation of the observed broadening and (partial) splitting of transitions in the fundamental band, and it is essential for understanding the intermolecular bending combination band where some transitions are completely split by isotope effects. The combination band is influenced by a significant bend-stretch Coriolis interaction and by the relatively large Xe isotope dependence of the intermolecular stretch frequency. The weak $CO_2$-Xe spectrum corresponding to the $(01^11) \leftarrow (01^10)$ hot band of $CO_2$ is also detected and analyzed, providing a measurement of the symmetry breaking of the $CO_2$ bending mode induced by the nearby Xe atom. This in-plane / out-of-plane splitting is determined to be 2.14 cm$^{-1}$.

### 1. Introduction

Rotationally resolved spectroscopy of weakly bound van der Waals complexes provides direct and specific information on intermolecular forces and dynamics. Dimers containing $CO_2$ and a rare gas (Rg) atom have been of notable experimental and theoretical interest, starting with the original observation of a microwave spectrum of $CO_2$-Ar by Steed et al. in 1979 [1]. Since then the majority of $CO_2$-Rg publications have continued to deal with $CO_2$-Ar, and for the particular case of $CO_2$-Xe there have been only three relevant experimental studies. First, in 1998 Randall et al. [2] measured the rotationally resolved infrared spectrum of $CO_2$-Xe in the region of the $CO_2$ $\nu_3$ fundamental band ($\approx$2350 cm$^{-1}$). Next, in 1993, Iida et al. [3] measured pure rotational microwave spectra of $CO_2$-Xe, and finally in 2006, Konno et al. [4] measured the $\nu_3$ region infrared spectrum of the $^{18}$O isotopologue of $CO_2$-Xe ($\approx$2314 cm$^{-1}$). There are six isotopes of xenon with significant natural abundances, as listed in Table 1. In their microwave spectra, Iida et al. [3] measured transitions involving the four most abundant of these ($^{129}$Xe, $^{131}$Xe, $^{132}$Xe, and $^{134}$Xe), but in the infrared spectra any isotopic splitting remained unresolved.

In the present paper, we revisit the spectrum of $CO_2$-Xe in the $CO_2$ $\nu_3$ region. Coverage of the $\nu_3$ fundamental band is expanded and the influence of the range of Xe isotopes is explored. The intermolecular bend and stretch modes are observed by means of their combination bands with the intramolecular $CO_2$ $\nu_3$ stretch, and here the Xe isotope effects turn out to be much larger than in the fundamental. Finally, $CO_2$-Xe is detected in the region of the $CO_2$ $(01^11) – (01^10)$ hot band near 2337 cm$^{-1}$, enabling observation of the splitting of the degenerate $CO_2$ bending vibration into two modes (in-plane and out-of-plane) due to the presence of the nearby Xe atom.

The minimum energy structure of $CO_2$-Xe (and the other $CO_2$-Rg dimers) is T-shaped, such that the Xe atom is located "beside" the linear $CO_2$ molecule with an effective C to Xe distance of about 3.8 Å [2]. The *a*-inertial axis of the dimer connects C and Xe, and the *A* rotational constant of $CO_2$-Xe is similar to the *B* constant of $CO_2$ ($\approx 0.39$ cm$^{-1}$). The *b*-axis is parallel to the O-C-O axis, and the *c*-axis is perpendicular to the $CO_2$-Xe plane. $CO_2$-Xe has six normal modes of vibration. Four of these correspond to intramolecular ($CO_2$) vibrations: the symmetric stretch ($\nu_1 \approx 1388$ cm$^{-1}$), the doubly-degenerate bend ($\nu_2 \approx 667$ cm$^{-1}$), and the asymmetric stretch ($\nu_3 \approx 2350$ cm$^{-1}$). The two remaining normal modes are intermolecular vibrations: the van der Waals stretch, and the bend. The T-shaped equilibrium structure corresponds to the $C_{2v}$ point group, for which fundamental vibrations can have A$_1$ symmetry ($\nu_1$ $CO_2$ stretch, in-plane component of the $\nu_2$ $CO_2$ bend, and van der Waals stretch), B$_2$ symmetry ($\nu_3$ $CO_2$ stretch, intermolecular bend), or B$_1$ symmetry (out-of-plane component of the $CO_2$ $\nu_2$ bend). A$_2$ symmetry modes can only occur for states which are combinations involving B$_1\times$B$_2$ modes. Note that the degeneracy of $\nu_2$ for the $CO_2$ monomer is lifted in $CO_2$-Xe, giving two distinct modes, with A$_1$ and B$_1$ symmetry. Nuclear spin statistics allow only even values of $K_a$ in A symmetry modes (including the ground state) of dimers containing C$^{16}$O$_2$, and only odd values in B symmetry modes. Two more recent theoretical publications have reported detailed potential energy surfaces for the $CO_2$-Xe interaction, using *ab initio* methods at the CCSD(T) level. The results were similar, with Chen and Zhu [5] obtaining a global minimum energy of -261.4 cm$^{-1}$ at an intermolecular distance of 3.81 Å, while Wang et al. [6] reported -258.8 cm$^{-1}$ at 3.78 Å. Notably, the former authors also included the dependence of the potential on the $CO_2$ $\nu_3$ vibration, allowing for more detailed comparisons with infrared spectra.

## 2. Results

### 2.1. Description of the spectra

Spectra were recorded using a pulsed supersonic slit jet expansion probed by a rapid-scan optical parametric oscillator source [7-9]. The gas expansion mixture contained about 0.03 % carbon dioxide plus 0.3 % xenon in helium carrier gas, and the backing pressure was about 12 atmospheres. Wavenumber calibration was made by simultaneous recording of signals from a fixed etalon and a reference gas cell containing room temperature $CO_2$. Spectral simulation and fitting were made using the PGOPHER software [10].

The observed spectra are shown in Figs. 1-4. The $\nu_3$ fundamental band in Fig. 1, which has been studied previously [2], and the $CO_2$ $(01^11) \leftarrow (01^10)$ hot bands in Fig. 4 have $b$-type rotational selection rules ($\Delta K_a = \pm 1$). These bands show relatively little effect of Xe isotope splitting (except at high $J$-values) because the intramolecular vibrational frequencies have almost no dependence on xenon atomic mass. The bending combination band in Figs. 2 and 3 has $a$-type selection rules ($\Delta K_a = 0$). It does show significant isotope splitting, especially in the $Q$- and $R$-branches (Fig. 3). The splitting arises mostly because there is a strong Coriolis interaction between the bend and the nearby intermolecular (van der Waals) stretch, whose frequency does have a significant dependence on xenon mass. In the following section, we first describe the approach we used to analyze the isotopic dependence of the spectrum. This approach was especially required for the combination band, but it also proved useful for the fundamental band.

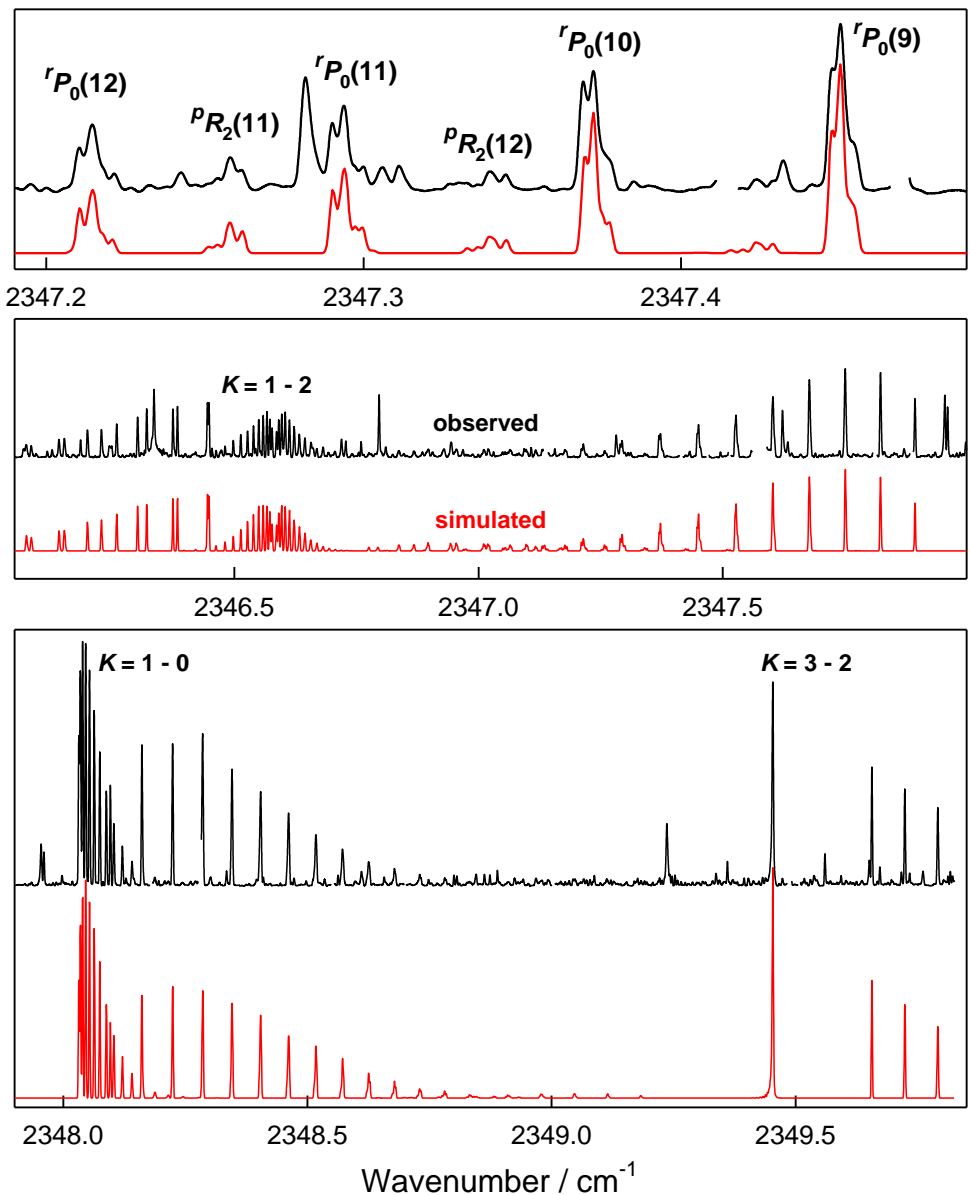

Fig. 1. Observed and simulated spectra of $CO_2$-Xe in the region of the $CO_2$ $\nu_3$ fundamental band. The simulation (red) represents the sum of the six Xe isotopes, with isotopic scaling as described in the text. Gaps in the observed spectrum correspond to regions of $CO_2$ monomer and $CO_2$-He absorption. Other observed lines not present in the simulation may be due to larger clusters such as $CO_2$-$Xe_2$. The center and bottom panels show an overview of the band, with the vertical scale magnified in the center panel by a factor of 1.5. The top panel is an expanded view showing the Xe isotope structure for higher $J$-value transitions, with the vertical scale magnified by a factor of 7.7 with respect to the bottom panel. Here each label refers to a single transition (e.g. $^{r}P_0(10)$) with semi-resolved Xe isotope structure.

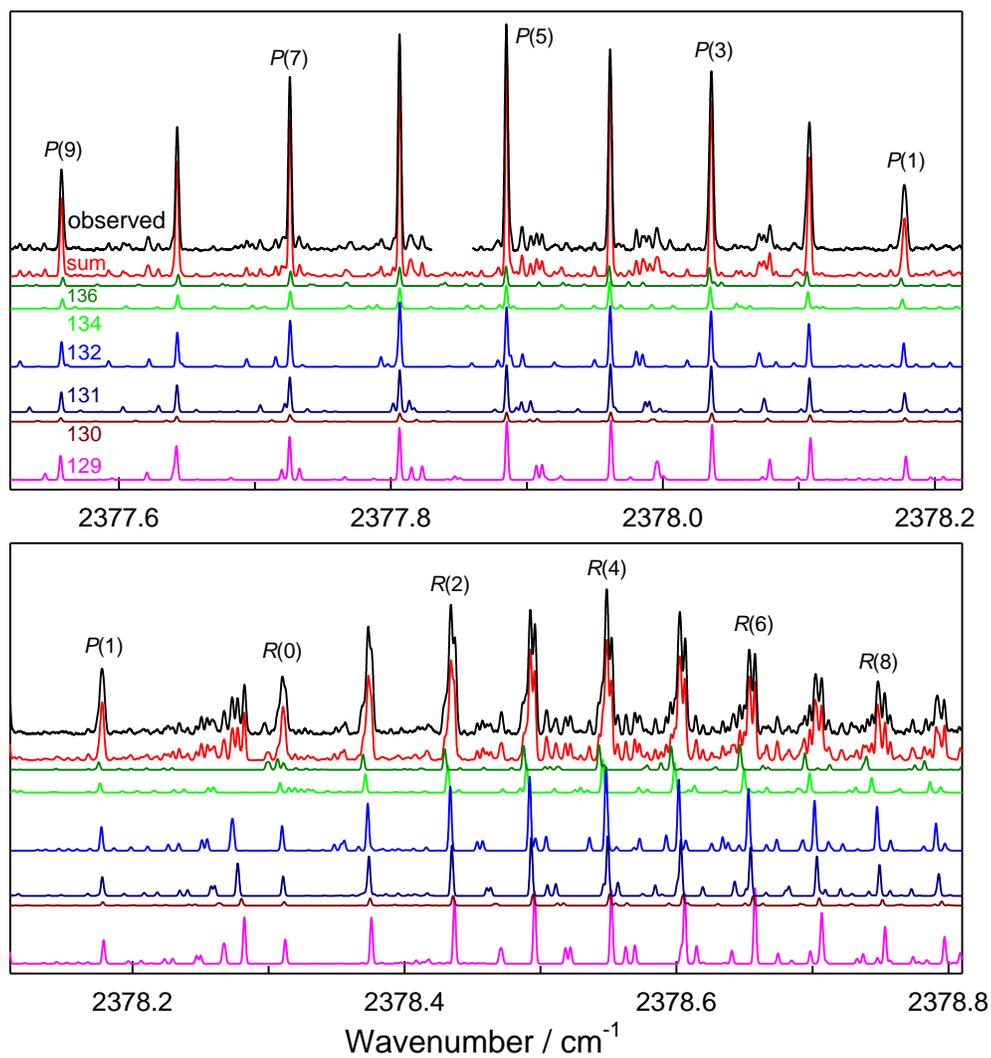

Fig. 2.   Observed and simulated spectra of $CO_2$-Xe in the region of the intermolecular bend combination band. Gaps in the observed spectrum correspond to regions of $CO_2$ monomer absorption. Simulated contributions of the various Xe isotopes are color-coded, with their sum shown in red.

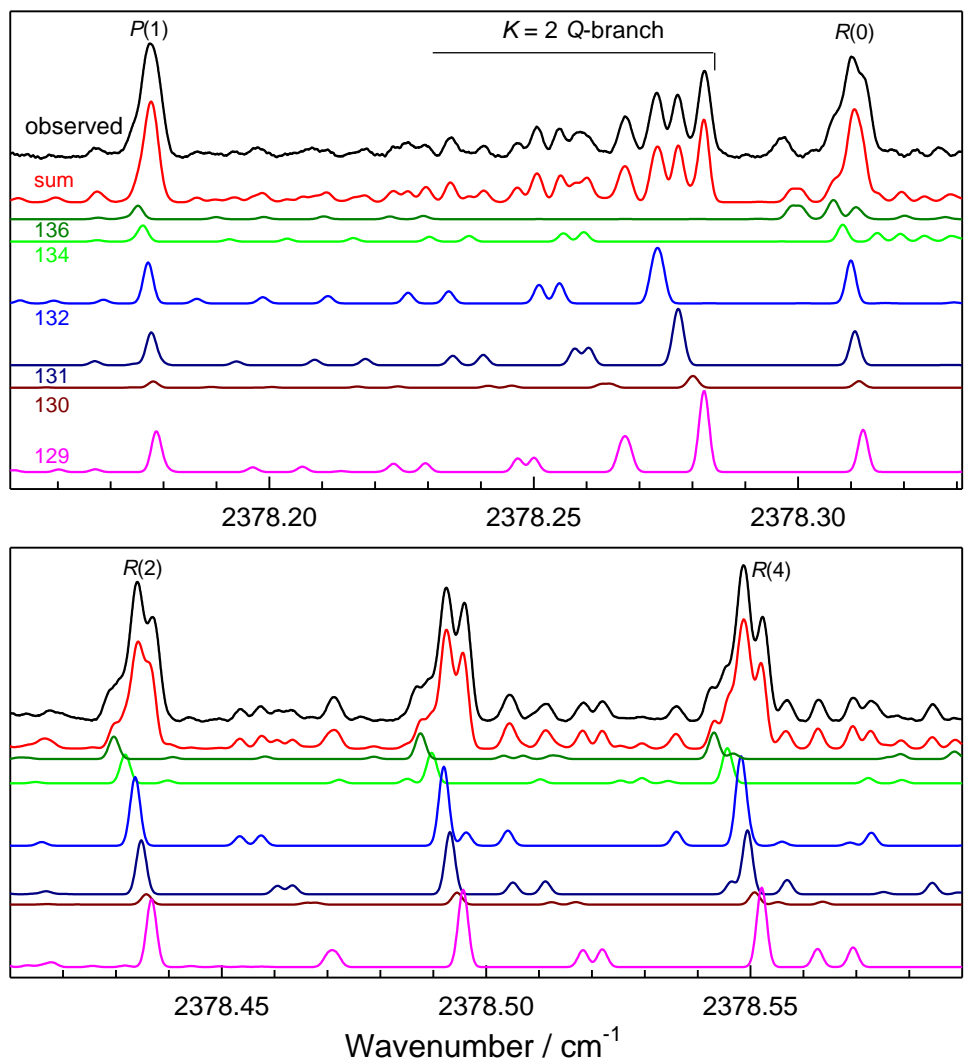

Fig. 3. Expanded view of the $Q$- and $R$-branch regions of the $CO_2$-Xe intermolecular bend combination band. Note the large isotopic splitting of some transitions and the good match between the observed spectrum (black) and the summed simulation (red) traces.

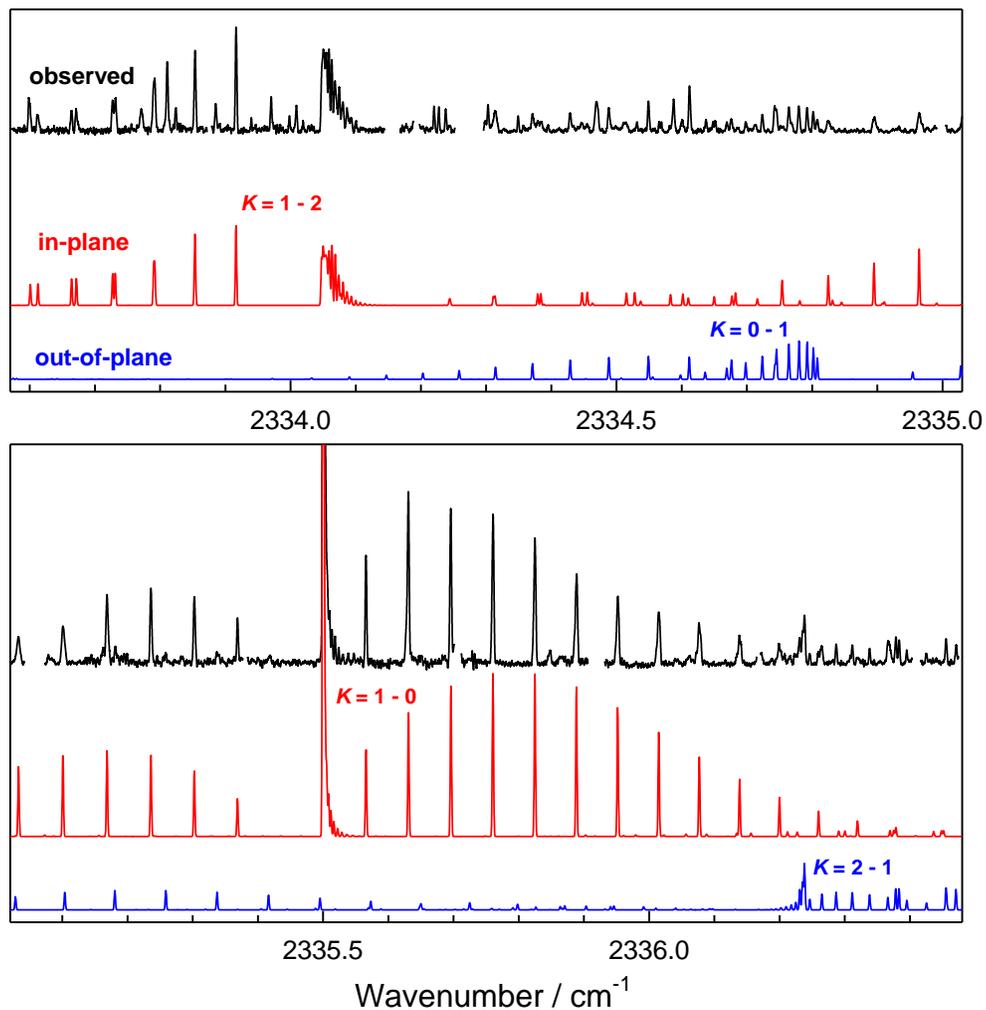

Fig. 4.   Observed and simulated spectra of $CO_2$-Xe in the region of the $CO_2$ $(01^11) \leftarrow (01^10)$ hot band. Gaps in the observed spectrum correspond to regions of $CO_2$ monomer absorption. The peak of the strong and mostly unresolved $K_a = 1 \leftarrow 0$ $Q$-branch at 2335.50 cm⁻¹ is truncated in order to display the rest of the spectrum more clearly.

### 2.2. Xenon isotope effects

The isotopes are far from being fully resolved in our infrared spectra, so we could not individually determine all the vibrational and rotational parameters for all six isotopes. Instead, we wanted to model the isotopic dependence using only a few free parameters. As a starting point, we determined the Xe isotopic dependence of the $B$ and $C$ rotational constants for a rigid T-shaped model of $CO_2$-Xe (the $A$ constant has no dependence in this model). Not surprisingly, $B$ simply varies as the ratio of the reduced mass of the $CO_2$ and Xe pair. $C$ has a somewhat different dependence, since the $CO_2$ monomer is not parallel to the $c$-axis.

Using these scaling factors, we tried fitting the 27 observed microwave transitions [3] of $CO_2$-[129]Xe, -[131]Xe, -[132]Xe, and -[134]Xe (with hyperfine splitting removed for [131]Xe) in a unified fashion using a single value for $B$ and $C$, with scaling, rather than individual values for each isotope. This gave good results, but there was still some remaining systematic dependence on xenon mass. This dependence was minimized by introducing new empirical parameters $B_{adj}$ and $C_{adj}$ which act as slight corrections to the rigid model scaling factors. The $B$ and $C$ rotational parameters are then given by

$$B(N) = B(N_0) \times [F_B(N) + B_{adj} \times (N - N_0)] \qquad (1)$$

$$C(N) = C(N_0) \times [F_C(N) + C_{adj} \times (N - N_0)], \qquad (2)$$

where N is the xenon atomic mass number (129, 130, etc.), $N_0$ is a "standard" atomic mass number (taken to be 131, closest to the weighted average atomic mass), and $F_B$ and $F_C$ are the rigid model scaling factors relative to [131]Xe, as listed in Table 1.

The microwave test fits also included the centrifugal distortion parameters $\Delta_{JK}$ and $\Delta_J$, which were scaled by the factors $F_B$ and $F_B{}^2$, respectively (without $B_{adj}$ or $C_{adj}$). In the original fit by Iida et al. [3], 27 microwave transitions (not counting hyperfine splitting) had been fitted using 12 adjustable parameters to obtain an rms deviation of 2 kHz. Our unified fit using 4 parameters ($B$, $C$, $\Delta_{JK}$, $\Delta_J$) gave a deviation of 200 kHz. By including the empirical parameters $B_{adj}$ and $C_{adj}$ this was reduced to 9 kHz, which we considered to be adequate for our infrared analysis. The advantage of the unified approach is that all the isotopes can be represented, including $^{130}$Xe and $^{136}$Xe for which microwave data are not available. Even better, the same rotational scaling factors can be used for excited state parameters in the infrared analysis, having been "verified" with the accurate ground state microwave data.

Table 1. Xenon isotope atomic mass, abundance, and rigid model scaling factors

| atomic mass number N | atomic mass (Dalton) | abundance | $F_B$ | $F_C$ |
|---|---|---|---|---|
| 129 | 128.905 | 0.264 | 1.003902 | 1.003576 |
| 130 | 129.904 | 0.041 | 1.001938 | 1.001776 |
| 131 | 130.905 | 0.212 | 1.000000 | 1.000000 |
| 132 | 131.904 | 0.269 | 0.998095 | 0.998253 |
| 134 | 133.905 | 0.104 | 0.994365 | 0.994834 |
| 136 | 135.907 | 0.089 | 0.990743 | 0.991507 |

To allow for isotopic variation of vibrational band origins, another empirical parameter called *Offset* was introduced, assuming a simple linear shift with isotope mass number, so that

$$\nu_0(N) = \nu_0(N_0) + \textit{Offset} \times (N - N_0). \tag{3}$$

The parameter *Offset* describes the band origin shift for each change of one in the Xe atomic mass number, so it has units of cm$^{-1}$ per Dalton. Of course we know that a harmonic stretching frequency actually scales as the square root of the appropriate reduced mass, and this is not quite linear in N. But in the present case the difference is essentially negligible.

### 2.3. The $\nu_3$ fundamental band

The value of the *A* rotational constant has only a very small effect on the microwave spectrum (where all transitions have $\Delta K_a = 0$) and must be determined using the infrared $\nu_3$ fundamental band (where $\Delta K_a = \pm 1$). So our microwave and fundamental band fit was a combined one, in which microwave measurements were weighted by a factor of 2000 to reflect their higher precision. In this fit, the excited state distortion parameters $\Delta_{JK}$ and $\Delta_J$ were constrained to equal their ground state values, as in much previous work on $CO_2$-Rg dimers [2]. Similarly, the parameters $B_{\text{adj}}$ and $C_{\text{adj}}$ were common to both states, as mentioned above. Input data for the fit consisted of the 27 microwave transitions, plus 89 infrared transitions, most of which were blends of two or more isotopes. Each blend was fitted to an intensity weighted average of its components using the Mergeblends option of PGOPHER [10].

The parameters resulting from the fit are listed in Table 2. As noted, these parameters apply specifically to $CO_2$-$^{131}$Xe, and they can also be considered as isotopic averages since $^{131}$Xe (atomic mass 130.905) is close to the weighted average mass (131.293). The scaled parameter values for the other isotopes, and the observed infrared transitions, are given as Supplementary Information. Average rms deviations in the fit were 9 kHz for the microwave measurements and 0.00020 cm$^{-1}$ for the infrared measurements. The latter is essentially equal to the experimental precision, while the experimental line width is about ten times greater. The simulated spectrum in Fig. 1 is based on the parameters in Table 2.

The empirical shift parameters $B_{adj}$ and $C_{adj}$ are based almost entirely on the isotopic microwave data, and are reasonably well determined. In contrast, the *Offset* parameter giving the isotopic dependence of the $\nu_3$ fundamental band origin is only marginally significant. Its magnitude implies a total difference of only about 0.0004 cm$^{-1}$ between $CO_2$-$^{136}$Xe and $CO_2$-$^{129}$Xe, and its negative sign means that the band origin shifts down slightly as the xenon atomic weight increases. This direction can be understood as an anharmonic effect due to the (slight) decrease in zero point intermolecular distance with increased mass. In other words, a heavier Xe atom is closer on average to $CO_2$ and the vibrational red shift increases. The actual value of this shift is -1.472 cm$^{-1}$ (for $^{131}$Xe) relative to the free $CO_2$ molecule, in good agreement with the value of -1.471 cm$^{-1}$ reported by Randall et al. [2]. The other parameters in Table 2 also agree well with Randall et al., but should be more reliable since we have a wider range of infrared data plus the advantage of including ground state microwave results [3].

The expanded view in the top panel in Fig. 2 shows how the Xe isotopic scaling in the combined fit gives a good representation of the broadening and (partially resolved) splitting of higher *J*-value transitions in the fundamental infrared spectrum. This is satisfying, but the real motivation for the scaling is to understand the combination bands where splittings are much larger as described in the following section.

Table 2. Molecular parameters for $CO_2$-Xe.[a]

| | $CO_2$-$^{131}$Xe Ground State | $CO_2$-$^{131}$Xe Fundamental State | $CO_2$-$^{131}$Xe Bend State | $CO_2$-$^{131}$Xe Stretch State |
|---|---|---|---|---|
| $\nu_0$ | | 2347.6713(1) | 2378.2452(1) | 2379.4514(3) [b] |
| $A$ | 0.395952(15) | 0.392803(21) | 0.407053(39) | [0.40] [b] |
| $B$ | 0.03534253(28) | 0.0353575(18) | 0.034146(36) | 0.034242(22) |
| $C$ | 0.03231320(28) | 0.0322985(13) | 0.031855(32) | 0.030814(13) |
| $10^6 \times \Delta_{JK}$ | 3.1084(47) | 3.1084 [c] | 14.7(20) | |
| $10^7 \times \Delta_J$ | 1.3270(31) | 1.3270 [c] | 2.73(40) | |
| $\xi_c$ | | | 0.029939(49) | |
| *Offset* | | -0.00006 (4) | -0.000645(28) | -0.02484(15) |
| $10^5 \times B_{\mathrm{adj}}$ | 1.83(17) | | | |
| $10^5 \times C_{\mathrm{adj}}$ | 0.84 (17) | | | |

[a] Units are cm$^{-1}$ except for the empirical parameters $B_{\mathrm{adj}}$, $C_{\mathrm{adj}}$ which are in (Dalton)$^{-1}$, and *Offset* which is in cm$^{-1}$/Dalton. (Dalton = atomic mass unit). Quantities in parentheses correspond to $1\sigma$ from the least-squares fit, in units of the last quoted digit.

[b] $A$ could not be well determined for the stretch state since only $K_a = 1$ levels were observed, and it was fixed at the indicated value. Thus the real uncertainty in the stretch band origin is limited by the possible uncertainty in $A$, about 0.01 cm$^{-1}$.

[c] These centrifugal distortion parameters were fixed at their ground state values in the combined microwave and $\nu_3$ fundamental band fit.

## 2.4. The bend and stretch combination bands

As we have just seen, the $\nu_3$ fundamental band origin has very little Xe isotope dependence, and isotope splitting is only partially evident for transitions with higher $J$-values where the isotopic dependence of the rotational constants becomes important. In contrast, it is clear by inspection that many transitions in the bending combination band (Figs. 2 and 3) show significant Xe isotopic splitting. The point here is that the $CO_2$-Xe intermolecular bending frequency is likely to be isotope dependent due to changes in the effective reduced mass for bending. But there is a more important factor, namely the proximity of the van der Waals stretch vibration, which is predicted [5,6] to lie within about 2 cm$^{-1}$ of the bend.

It turns out that the stretch lies 1.24 cm$^{-1}$ above the bend, and the Coriolis interaction between bend and stretch has a large effect on the bending combination band, even though it appears that the stretch combination band has little or no strength on its own. The $c$-type Coriolis interaction is characterized by the matrix element $\langle$bend, $J$, $k$ $|H$ $|$stretch, $J$, $k \pm 1\rangle = \frac{1}{2} \xi_c \times [J(J + 1) - k(k \pm 1)]^{\frac{1}{2}}$, where $k$ is signed $K_a$, and $\xi_c$ is the Coriolis interaction parameter. The influence of the stretch is accentuated by the fact that the stretch frequency has a relatively large Xe isotope dependence, and by the fact that the $K_a = 2$ levels of the bend happen to lie very close to the $K_a = 1$ levels of the stretch. As mentioned above, the bend combination band has $a$-type selection rules ($\Delta K_a = 0$), and its most prominent transitions are the $P$- and $R$-branches with $K_a = 0$, as labeled in Fig. 2. Note how the $P$-branch lines remain relatively sharp even for higher $J$-values, while the $R$-branch lines show more isotopic splitting, reducing their peak height. The difference arises because vibrational and rotational isotopic shifts tend to cancel in the $P$-branch and to add in the $R$-branch. Isotope splitting is especially noticeable for $Q$-branch transitions with $K_a = 2$, shown in detail in the upper panel of Fig. 3. Note that the $K_a = 0$ $Q$-branch is forbidden, and the $K_a = 4$ $Q$- branch (at about 2378.41 cm$^{-1}$) is very weak since there is little $K_a = 4$ population at the experimental temperature of $\approx 2$ K.

Our fit of the bend combination band involved a total of 125 observed lines which were fitted with an rms average error of 0.00034 cm$^{-1}$. As can be seen from the simulated spectra in Fig. 3, many of these lines show partial or complete isotopic resolution, while many others are blends. A significant number of the lines were actually assigned to the stretch combination band, where transitions to excited state $K_a = 1$ levels borrow intensity due to their mixing with $K_a = 2$ of the bend. In the fit, ground state parameters were fixed at the values already determined, and upper state rotational parameters and band origins were isotopically scaled as described previously. The $c$-type Coriolis parameter was scaled the same as the $C$ constant. There were 12 adjustable parameters, whose fitted values are shown in the right-hand columns of Table 2. For the bending state these parameters were $\nu_0$, $A$, $B$, $C$, $\Delta_{JK}$, and $\Delta_J$. For the stretching state they were $\nu_0$, $B$, and $C$, but $A$ could not be well determined. In addition, the Coriolis parameter $\xi_c$ was determined, as were the *Offset* parameters describing the Xe isotope scaling of the two band origins. The observed bend-stretch Coriolis interaction parameter, $\xi_c = 0.0299$ cm$^{-1}$, implies a value of about 0.48 for the dimensionless Coriolis zeta parameter, $\zeta_c$.

By subtracting the $\nu_3$ fundamental band origin from the combination band origins in Table 2 we determine experimental intermolecular bend and stretch frequencies for $CO_2$-Xe (in the $CO_2$ $\nu_3$ state). These values are listed in Table 3, where they are compared with previous estimates from *ab initio* theory [5,6] and from an experimentally based empirical harmonic force field [3]. The agreement with our experimental results is quite satisfactory, particularly for the calculations of Chen and Zhu [5]. As mentioned, the $A$ rotational constant could not be determined for the stretching combination state, since only $K_a = 1$ levels were observed. As a result, the real uncertainty in the stretching combination band origin (Table 2) and the derived stretching frequency (Table 3) are perhaps around 0.01 cm$^{-1}$. In other words, even though $K_a = 1$ levels of the stretch state were precisely (0.002 cm$^{-1}$) determined, the (forbidden) $K_a = 0$ levels which define the origin still have an uncertainty of around 0.01 cm$^{-1}$.

Table 3. Comparison of experimental and theoretical intermolecular vibrational frequencies for $CO_2$-Xe (in cm$^{-1}$).

| | Experiment $CO_2$ (001) present work [a] | | Theory $CO_2$ (001) Ref. 5 | Theory $CO_2$ (000) Ref. 5 | Theory $CO_2$ (000) Ref. 6 | Empirical $CO_2$ (000) Ref. 3 |
|---|---|---|---|---|---|---|
| | $^{131}$Xe | $^{129}$Xe − $^{136}$Xe | | | | |
| Bend | 30.574 | 30.576 − 30.571 | 30.441 | 30.782 | 29.021 | 32.8 |
| Stretch | 31.780 | 31.830 − 31.656 | 31.151 | 31.020 | 31.177 | 33.2 |

[a]For the bend, the experimental uncertainty is less than 0.001 cm$^{-1}$. For the stretch, this uncertainty is larger (≈0.01 cm$^{-1}$) because of uncertainty in the exact $A$ value in the excited stretch state (see text).

The scaling parameters (called *Offset*) for the bending and stretching vibrations describe how much each band origin shifts for a change of one Dalton in the Xe atomic mass. Although *Offset* for the bend is ten times larger than that determined above for the $\nu_3$ fundamental band, it is still quite small, just -0.0006 cm$^{-1}$ per Dalton. But *Offset* for the stretch is -0.025 cm$^{-1}$ per Dalton, a further 40 times greater in magnitude. The negative signs show that the bend and stretch frequencies decrease with increasing Xe mass, as expected. Interestingly, the observed value of the stretch *Offset* parameter corresponds to that expected for a harmonic frequency of 26.2 cm$^{-1}$, while the actual stretch frequency (Table 3) is 31.78 cm$^{-1}$. Thus the Xe isotope dependence of the $CO_2$-Xe intermolecular stretch frequency is somewhat smaller than might be expected.

Figure 5 illustrates the interaction between $K_a = 2$ of the bend and $K_a = 1$ of the stretch, explaining why the isotopic shifts are so prominent for these levels. Here the actual (perturbed) energy levels for $J = 2$ are shown as red and blue circles, and the deperturbed levels (that is, with Coriolis interaction removed) are smaller black circles and lines. For $CO_2$-$^{129}$Xe, the stretch levels are above the bend levels by more than 0.1 cm$^{-1}$. But with increasing Xe atomic mass, the stretching frequency declines much faster than the bending frequency, so that for $^{134}$Xe the unperturbed levels are virtually coincident and then for $^{136}$Xe they cross so that the stretch is below the bend. Of course in reality the crossing is avoided due to the Coriolis interaction, so the real bend and stretch levels never become any closer than about 0.06 cm$^{-1}$ (twice the Coriolis matrix element).

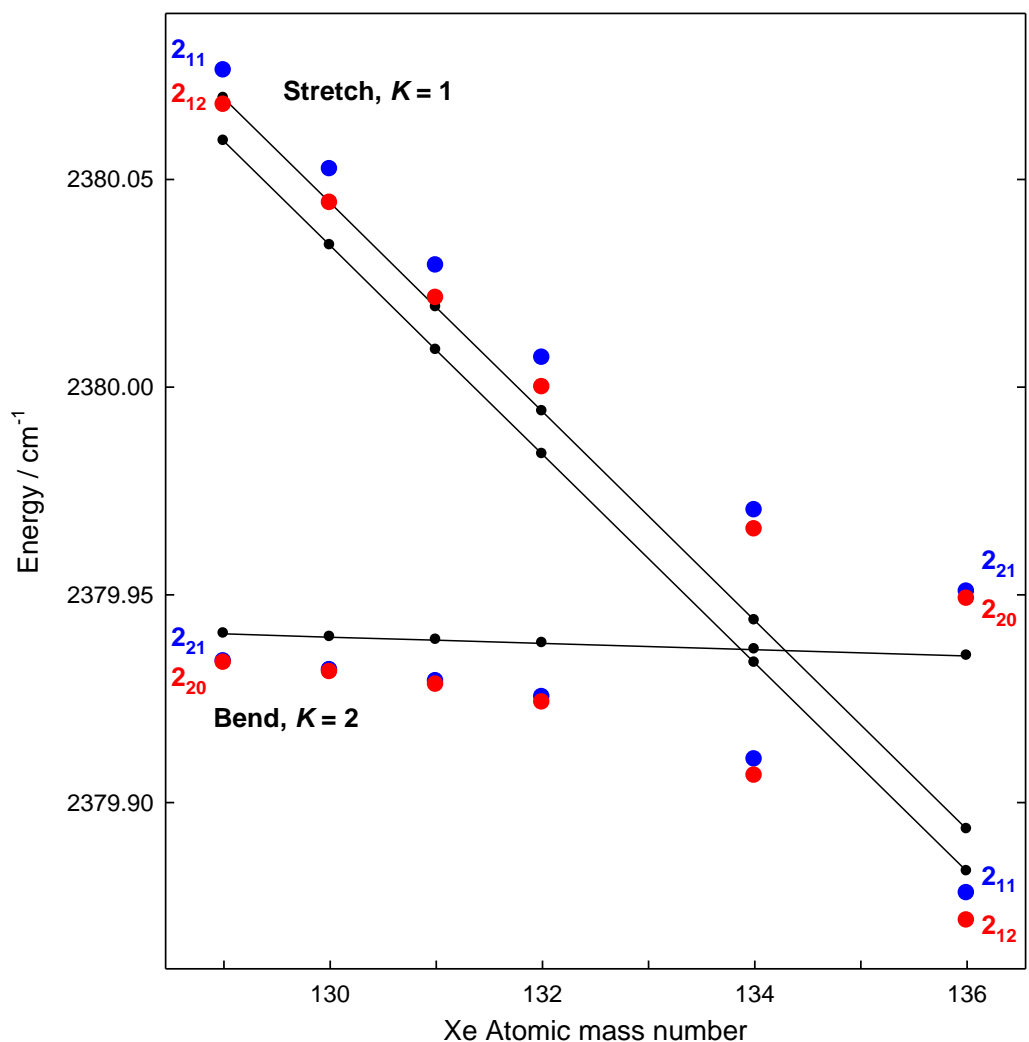

Fig. 5. Energy level diagram illustrating the Xe isotope dependence of the Coriolis interaction between the intermolecular bend and stretch states of $CO_2$-Xe for $J = 2$. The actual (perturbed) energies of the $J_{KaKc} = 2_{20}$, $2_{21}$, $2_{11}$, and $2_{12}$ rotational levels are shown by red and blue circles. Their deperturbed energies (Coriolis interaction removed) are shown as small black circles connected by lines (the asymmetry splitting between $2_{20}$ and $2_{21}$ is too small to see here). For $CO_2$-$^{129}$Xe, the $K_a = 1$ levels of the stretch lies well above $K_a = 2$ of the bend. But the stretching frequency decreases with increasing Xe atomic mass, so that the deperturbed bend and stretch levels are almost coincident for $CO_2$-$^{134}$Xe, and the stretch lies below the bend for $CO_2$-$^{136}$Xe.

### 2.5. The $CO_2$ $(01^11) - (01^10)$ hot bands

In our supersonic expansion, there is sufficient population of $CO_2$-Xe in the excited $CO_2$ $\nu_2$ intramolecular bending state [$(\nu_1, \nu_2^{l2}, \nu_3) = (01^10)$ at $\approx 667$ cm$^{-1}$] to enable observation of the $(01^11) \leftarrow (01^10)$ hot band near 2337 cm$^{-1}$ in the $\nu_3$ region. The situation is similar to that described for $CO_2$-Ar in a recent paper, to which the reader is referred for more details [11]. Analogous spectra have also been observed for $C_2H_2$-Ar [12] and $CO_2$-$N_2$ [13]. The nearby Xe atom splits the degenerate $CO_2$ bend into an in-plane (i-p) and an out-of-plane (o-p) mode. In the $(01^10)$ lower state, these modes have $A_1$ or $B_1$ symmetry, and hence allowed $K_a$ = even or odd levels, respectively. In the $(01^11)$ upper state, the symmetries become $B_2$ or $A_2$ with allowed $K_a$ = odd or even levels, respectively. The $(01^11) \leftarrow (01^10)$ hot band has $b$-type selection rules ($\Delta K_a = \pm 1$) just like the $\nu_3$ fundamental. The i-p component of the hot band has subbands with $K_a$ = odd $\leftarrow$ even, and the o-p component has $K_a$ = even $\leftarrow$ odd. There is strong $b$-type Coriolis mixing between the i-p and o-p modes, characterized by the matrix element $\langle$i-p, $J$, $k|H|$o-p, $J$, $k \pm 1\rangle = $½ $\xi_b \times [J(J + 1) - k(k \pm 1)]^{1/2}$, where $k$ is signed $K_a$, and $\xi_b$ is the Coriolis parameter. As for the intermolecular Coriolis interaction in the previous section, the rotational constant and the dimensionless zeta parameter $\zeta_b$ are related by $\xi_b = 2B\zeta_b$.

By analogy with $CO_2$-Ar [11] we expected the $CO_2$-Xe rotational constants and hot band vibrational shift to be similar to those of the fundamental, and also expected $\zeta_b \approx 1$. So to assign the spectrum (Fig. 4), the main unknown quantity was the splitting between the i-p and o-p modes, expected to be similar in the lower and upper states. After some trial and error, we found that this splitting had a value of about 2.1 cm$^{-1}$ (compared to 0.88 cm$^{-1}$ for $CO_2$-Ar) and assigned a total of 74 hot band lines. The detailed assignments are given as Supplementary Information. There was little isotope splitting, so this was ignored in the fit, meaning that the result should approximate that for the weighted average atomic mass (131.29). The rms error in the fit was 0.00028 cm$^{-1}$ and the parameters resulting from the fit are listed in Table 4. We did not attempt to determine centrifugal distortion parameters, but found that the fit was improved slightly by fixing $\Delta_{JK}$ and $\Delta_J$ at their ground state values from Table 2.

Table 4. Molecular parameters for the $(01^11) \leftarrow (01^10)$ hot band of $CO_2 - Xe$ (in cm$^{-1}$).[a]

|  | (010) i-p | (010) o-p | (011) i-p | (011) o-p |
|---|---|---|---|---|
| $\sigma_0$ | X[b] | 2.1404(43)+X | 2335.1405(5)+X | 2337.3115(43)+X |
| $A$ | 0.396555(30) | 0.395905(40) | [0.393455][c] | 0.392830(43) |
| $B$ | 0.035197(84) | 0.035467(17) | 0.035309(15) | 0.035447(35) |
| $C$ | 0.032480(79) | 0.0323282(81) | 0.0323751(55) | 0.032325(19) |
| $10^6 \times \Delta_{JK}$ | [3.1] | [3.1] | [3.1] | [3.1] |
| $10^7 \times \Delta_J$ | [1.3] | [1.3] | [1.3] | [1.3] |
| $\xi_b$ | 0.06999(25) | | [0.06999][d] | |

[a] Quantities in parentheses correspond to 1σ from the least-squares fit, in units of the last quoted digit. The parameters $\Delta_{JK}$ and $\Delta_J$ were fixed at the indicated ground state values.

[b] X is equal to the free $CO_2$ $\nu_2$ frequency (667.380 cm$^{-1}$) plus or minus an unknown vibrational shift which is unlikely to be more than a few cm$^{-1}$.

[c] $A' - A''$ was constrained to equal -0.0031 cm$^{-1}$.

[d] $\xi_b$ was constrained to be equal in the upper and lower states.

By far the strongest feature in the spectrum is the narrow $Q$-branch of the $K_a = 1 \leftarrow 0$ subband of the i-p mode at 2335.500 cm$^{-1}$ (see Fig. 4). Also labeled in Fig. 4 are the $K_a = 1 \leftarrow 2, 0$

← 1, and 2 ← 1 subbands. In addition, two very weak lines with $K_a = 3 \leftarrow 2$ were assigned (not shown in the figure). In the case of $CO_2$-Ar, some cross transitions (i-p ← o-p and o-p ←i-p) were observed [11]. However, such "forbidden" transitions could not be detected in the present case. They are much weaker for $CO_2$-Xe because the wider spacing of the i-p and o-p modes reduces Coriolis mixing. It was not possible to reliably determine all of the parameters in Table 4 at the same time without some constraint. After trying various schemes, we decided on the following two constraints: $(A' - A'') = -0.0031$ cm$^{-1}$ for the i-p state (similar to the $\nu_3$ fundamental band), and $\xi_b' = \xi_b''$ for the Coriolis parameter. Both constraints are reasonable based on the results for $CO_2$-Ar, where more parameters could be varied thanks to more data [11].

The $\xi_b$ parameter from Table 4 results in a value of about 0.990 for $\zeta_b$, depending slightly on exactly what value is used for $B$. This can be compared to reported values of 1.01 for $CO_2$-Ar and 1.003 for $CO_2$-$N_2$ [11,13]. These near unity values of $\zeta_b$ are a reflection of the fact that the $CO_2$ bend is only slightly perturbed in these weakly-bound complexes. An important aspect of the results is the splitting between i-p and o-p modes, determined to be 2.140(4) cm$^{-1}$ for $CO_2$-Xe in the lower $(01^10)$ state and 2.171(4) cm$^{-1}$ in the upper $(01^11)$ state, with o-p higher than i-p. This is larger than the 0.877 cm$^{-1}$ splitting [11] observed in $CO_2$-Ar and quite similar to the 2.307 cm$^{-1}$ splitting [13] in $CO_2$-$N_2$. Although the splittings are determined by our hot band spectrum, the actual $CO_2$ bending frequency is not. The parameter X in Table 4 represents this bending frequency (for the i-p mode), and X is equal to 667.38 cm$^{-1}$ plus or minus an unknown (but relatively small) vibrational shift. The $(01^11) \leftarrow (01^10)$ hot band origin for the free $CO_2$ molecule [14] is 2336.633 cm$^{-1}$, and the present $CO_2$-Xe origins from Table 4 represent vibrational shifts of -1.493 and -1.462 cm$^{-1}$ for the i-p and o-p modes, respectively. These are very similar to the shift of -1.472 cm$^{-1}$ as determined above for the $\nu_3$ fundamental band.

The strongest lines in the $CO_2$-Xe $(01^11) \leftarrow (01^10)$ hot band were about 0.02 times the strength of those in the $(001) \leftarrow (000)$ fundamental band. This is similar to $CO_2$-Ar, so we can repeat the conclusions made previously [11]. Very roughly, the observed hot band strength is about half of that expected if all the original room temperature population of $CO_2$ in the $(01^10)$ state survived the supersonic expansion without any relaxation. This relatively inefficient vibrational relaxation is probably related to the fact that the gas mixture contained over 99% helium in both cases.

### 3. Conclusions

In this paper, we have shown how the Xe isotope dependence of the $CO_2$-Xe spectrum can be parameterized in order to allow fitting and simulation of infrared spectra, even though the isotopic splittings range from unresolved through partly resolved (e.g. for higher-$J$ lines of the $\nu_3$ fundamental band) to fully resolved (e.g. for $K_a = 2$ lines of the bending combination band). Implementation of this parameterized fitting depended on the availability of precise, though incomplete, microwave data [3], and was greatly facilitated by the various features of the PGOPHER software package [10] which allow parameters to be varied and/or constrained in a flexible and user-friendly fashion. The most significant new results obtained here are determination of the intermolecular bend and intermolecular stretch frequencies for $CO_2$-Xe in the $CO_2$ $\nu_3$ state (and their Xe isotope dependence), and characterization of the splitting of the degenerate $CO_2$ $\nu_2$ bending vibration into in-plane and out-of-plane modes.

In conclusion, new infrared spectra of the weakly-bound $CO_2$-Xe complex are reported and their Xe isotope dependence is investigated in detail. The fundamental band ($CO_2$ $\nu_3$) is analyzed together with previous microwave data in terms of a unified set of parameters including Xe isotope scaling. Infrared combination bands involving the intermolecular bend and (van der Waals) stretch are observed for the first time, yielding experimental frequencies of 30.57 and 31.78 cm$^{-1}$, respectively (in the excited $CO_2$ $\nu_3$ state). Isotope splitting in the combination bands is accentuated by the strong bend-stretch Coriolis interaction and the relatively large isotope dependence of the stretch frequency. $CO_2$-Xe spectra are also observed corresponding to the $CO_2$ $(01^11) \leftarrow (01^10)$ hot band, and their analysis results in a value of 2.14 cm$^{-1}$ for the splitting of the $CO_2$ $\nu_2$ bend into in-plane and out-of-plane modes, strongly linked by Coriolis interaction.

**Supplementary Material**

Supplementary Material includes tables giving observed and fitted line positions for $CO_2$-Xe and detailed isotopic dependence of molecular parameters.

**Acknowledgements**

The financial support of the Natural Sciences and Engineering Research Council of Canada is gratefully acknowledged.

**References**

1 J.M. Steed, T.A. Dixon, and W. Klemperer, J. Chem. Phys. **70**, 4095-4100 (1979); Erratum: **75**, 5977 (1981).

2 R.W. Randall, M.A. Walsh, and B.J. Howard, Faraday Discuss. Chem. Soc. **85**, 13-21 (1988).

3 M. Iida, Y. Ohshima, and Y. Endo, J. Phys. Chem. **97**, 357-362 (1993).

4 T. Konno, S. Fukuda, and Y. Ozaki, Chem. Phys. Lett. **421**, 421-426 (2006).

5 M. Chen and H. Zhu, J. Theo. Comp. Chem. **11**, 537-546 (2012).

6 Z. Wang, E. Feng, C. Zhang, and C. Sun, Chem. Phys. Lett. **619**, 14-17 (2015).

7 M. Dehghany, M. Afshari, Z. Abusara, C. Van Eck, and N. Moazzen-Ahmadi, J. Mol. Spectrosc. **247**, 123-127 (2008).

8 N. Moazzen-Ahmadi and A.R.W. McKellar, Int. Rev. Phys. Chem. **32**, 611-650 (2013).

9 M. Rezaei, S. Sheybani-Deloui, N. Moazzen-Ahmadi, K.H. Michaelian, and A.R.W. McKellar, J. Phys. Chem. A, **117**, 9612-9620 (2013).

10 C. M. Western, *J. Quant. Spectrosc Rad. Transfer* **186**, 221 (2017); PGOPHER version 10.1, C. M. Western, 2018, University of Bristol Research Data Repository, doi:10.5523/bris.3mqfb4glgkr8a2rev7f73t300c

11 T.A. Gartner, A.J. Barclay, A.R.W. McKellar, and N. Moazzen-Ahmadi, Phys. Chem. Chem. Phys. **22**, 21488-21493 (2020).

12 Y. Ohshima, Y. Matsumoto, M. Takami, and K. Kuchitsu, J. Chem. Phys. **99**, 8385-8397 (1993).

13 A.J. Barclay, A.R.W. McKellar, and N. Moazzen-Ahmadi, J. Chem. Phys. **153**, 014303 (2020).

14 D. Bailly, R. Farrenq, G. Guelachvili, and C. Rosetti, J. Mol. Spectrosc. **90**, 74-105 (1981).

Appendix to:

# New infrared spectra of $CO_2 – Xe$: modeling Xe isotope effects, intermolecular bend and stretch, and symmetry breaking of the $CO_2$ bend

Table A-1. Molecular parameters for $CO_2$-Xe, microwave and fundamental band (in $cm^{-1}$). See also Table 2 of the paper.

|  | $CO_2$-$^{129}$Xe | $CO_2$-$^{130}$Xe | $CO_2$-$^{131}$Xe | $CO_2$-$^{132}$Xe | $CO_2$-$^{134}$Xe | $CO_2$-$^{136}$Xe |
|---|---|---|---|---|---|---|
| $\nu_0$ | 2347.6714 | 2347.6713 | 2347.6713(1) | 2347.6712 | 2347.6711 | 2347.6710 |
| $A'$ | 0.392803 | 0.392803 | 0.392803(21) | 0.392803 | 0.392803 | 0.392803 |
| $B'$ | 0.0354942 | 0.0354254 | 0.0353574(18) | 0.0352908 | 0.0351602 | 0.0350334 |
| $C'$ | 0.0324135 | 0.0323556 | 0.0322986(13) | 0.0322424 | 0.0321325 | 0.0320256 |
| $A''$ | 0.395952 | 0.395952 | 0.395952(15) | 0.395952 | 0.395952 | 0.395952 |
| $B''$ | 0.03547915 | 0.03541038 | 0.03534253(28) | 0.03527585 | 0.03514531 | 0.03501859 |
| $C''$ | 0.03242821 | 0.03237032 | 0.03231321(28) | 0.03225702 | 0.03214709 | 0.03204012 |
| $10^6 \times \Delta_{JK}$ | 3.1206 | 3.1145 | 3.1084(47) | 3.1025 | 3.0909 | 3.0797 |
| $10^7 \times \Delta_J$ | 1.3374 | 1.3322 | 1.3270(31) | 1.3220 | 1.3121 | 1.3026 |

Table A-2. Molecular parameters for $CO_2$-Xe, combination bands (in cm$^{-1}$). See also Table 2 of the paper.

| | $CO_2$-$^{129}$Xe | $CO_2$-$^{130}$Xe | $CO_2$-$^{131}$Xe | $CO_2$-$^{132}$Xe | $CO_2$-$^{134}$Xe | $CO_2$-$^{136}$Xe |
|---|---|---|---|---|---|---|
| Bend | | | | | | |
| $\nu_0$ | 2378.2465 | 2378.2459 | 2378.2452(1) | 2378.2446 | 2378.2433 | 2378.2420 |
| $A'$ | 0.407053 | 0.407053 | 0.407053(39) | 0.407053 | 0.407053 | 0.407053 |
| $B'$ | 0.034278 | 0.034212 | 0.034146(36) | 0.034082 | 0.033955 | 0.033833 |
| $C'$ | 0.031969 | 0.031911 | 0.031855(32) | 0.031800 | 0.031691 | 0.031586 |
| $10^6 \times \Delta_{JK}$ | 14.7 | 14.7 | 14.7(20) | 14.6 | 14.6 | 14.5 |
| $10^7 \times \Delta_J$ | 2.75 | 2.74 | 2.73(40) | 2.72 | 2.70 | 2.68 |
| $\xi_c$ | 0.030045 | 0.029992 | 0.029939(49) | 0.029887 | 0.029785 | 0.029686 |
| Stretch | | | | | | |
| $\nu_0$ | 2379.5011 | 2379.4763 | 2379.4514(3) | 2379.4266 | 2379.3769 | 2379.3272 |
| $A'$ | [0.40] | [0.40] | [0.40] | [0.40] | [0.40] | [0.40] |
| $B'$ | 0.034374 | 0.034307 | 0.034242(22) | 0.034177 | 0.034051 | 0.033928 |
| $C'$ | 0.030924 | 0.030869 | 0.030814(13) | 0.030760 | 0.030656 | 0.030554 |

Table A-3. Observed and fitted microwave transitions of $CO_2$-Xe (MHz).

Observed transitions from: M. Iida, Y. Ohshima, and Y. Endo, J. Phys. Chem. **97**, 357-362 (1993).

Calculated transitions from the combined microwave and fundamental band fit described in the paper.

| Isotope | $J'$ | $Ka'$ | $Kc'$ | $J''$ | $Ka''$ | $Kc''$ | Observed | Calculated | Obs - Calc |
|---|---|---|---|---|---|---|---|---|---|
| CO2-Xe-129 | 3 | 0 | 3 | 2 | 0 | 2 | 6104.700 | 6104.689 | 0.011 |
| CO2-Xe-129 | 3 | 2 | 1 | 2 | 2 | 0 | 6107.053 | 6107.069 | -0.016 |
| CO2-Xe-129 | 4 | 0 | 4 | 3 | 0 | 3 | 8136.446 | 8136.438 | 0.008 |
| CO2-Xe-129 | 4 | 2 | 2 | 3 | 2 | 1 | 8144.546 | 8144.557 | -0.011 |
| CO2-Xe-129 | 4 | 2 | 3 | 3 | 2 | 2 | 8138.771 | 8138.776 | -0.005 |
| CO2-Xe-129 | 5 | 0 | 5 | 4 | 0 | 4 | 10165.494 | 10165.493 | 0.001 |
| CO2-Xe-129 | 5 | 2 | 3 | 4 | 2 | 2 | 10183.585 | 10183.585 | 0.000 |
| CO2-Xe-129 | 5 | 2 | 4 | 4 | 2 | 3 | 10172.026 | 10172.026 | 0.000 |
| CO2-Xe-129 | 6 | 0 | 6 | 5 | 0 | 5 | 12191.179 | 12191.186 | -0.007 |
| CO2-Xe-129 | 7 | 0 | 7 | 6 | 0 | 6 | 14212.849 | 14212.859 | -0.010 |
| CO2-Xe-131 | 3 | 0 | 3 | 2 | 0 | 2 | 6082.105 | 6082.095 | 0.010 |
| CO2-Xe-131 | 4 | 0 | 4 | 3 | 0 | 3 | 8106.363 | 8106.356 | 0.007 |
| CO2-Xe-131 | 5 | 0 | 5 | 4 | 0 | 4 | 10127.961 | 10127.958 | 0.003 |
| CO2-Xe-131 | 6 | 0 | 6 | 5 | 0 | 5 | 12146.246 | 12146.243 | 0.003 |
| CO2-Xe-132 | 3 | 0 | 3 | 2 | 0 | 2 | 6071.071 | 6071.063 | 0.008 |
| CO2-Xe-132 | 3 | 2 | 1 | 2 | 2 | 0 | 6073.351 | 6073.357 | -0.006 |
| CO2-Xe-132 | 4 | 0 | 4 | 3 | 0 | 3 | 8091.674 | 8091.666 | 0.008 |
| CO2-Xe-132 | 4 | 2 | 2 | 3 | 2 | 1 | 8099.565 | 8099.565 | 0.000 |
| CO2-Xe-132 | 4 | 2 | 3 | 3 | 2 | 2 | 8093.911 | 8093.907 | 0.004 |
| CO2-Xe-132 | 5 | 0 | 5 | 4 | 0 | 4 | 10109.632 | 10109.629 | 0.003 |
| CO2-Xe-132 | 5 | 2 | 4 | 4 | 2 | 3 | 10115.986 | 10115.963 | 0.023 |
| CO2-Xe-132 | 5 | 2 | 3 | 4 | 2 | 2 | 10127.275 | 10127.275 | 0.000 |
| CO2-Xe-132 | 6 | 0 | 6 | 5 | 0 | 5 | 12124.296 | 12124.297 | -0.001 |
| CO2-Xe-132 | 7 | 0 | 7 | 6 | 0 | 6 | 14135.036 | 14135.025 | 0.011 |
| CO2-Xe-134 | 3 | 0 | 3 | 2 | 0 | 2 | 6049.461 | 6049.470 | -0.009 |
| CO2-Xe-134 | 4 | 0 | 4 | 3 | 0 | 3 | 8062.900 | 8062.916 | -0.016 |
| CO2-Xe-134 | 5 | 0 | 5 | 4 | 0 | 4 | 10073.736 | 10073.755 | -0.019 |

Table A-3. Observed and fitted fundamental band transitions of CO$_2$-Xe (cm$^{-1}$).

Int = calculated relative intensity, used to weight the blended lines.

| Isotope | $J'$ | $Ka'$ | $Kc'$ | $J''$ | $Ka''$ | $Kc''$ | Observed | Calculated | Obs - Calc | Int |
|---------|------|-------|-------|-------|--------|--------|----------|------------|-----------|-----|
| CO2-Xe-132 | 5 | 3 | 2 | 6 | 4 | 3 | 2344.7035 | 2344.7037 | -0.0003 | 0.00187 |
| CO2-Xe-132 | 5 | 3 | 3 | 6 | 4 | 2 | 2344.7035 | 2344.7037 | -0.0003 | 0.00187 |
| CO2-Xe-129 | 5 | 3 | 2 | 6 | 4 | 3 | 2344.7035 | 2344.7030 | 0.0005 | 0.00183 |
| CO2-Xe-129 | 5 | 3 | 3 | 6 | 4 | 2 | 2344.7035 | 2344.7030 | 0.0005 | 0.00183 |
| CO2-Xe-131 | 5 | 3 | 2 | 6 | 4 | 3 | 2344.7035 | 2344.7035 | 0.0000 | 0.00147 |
| CO2-Xe-131 | 5 | 3 | 3 | 6 | 4 | 2 | 2344.7035 | 2344.7035 | 0.0000 | 0.00147 |
| CO2-Xe-134 | 5 | 3 | 2 | 6 | 4 | 3 | 2344.7035 | 2344.7042 | -0.0007 | 0.00073 |
| CO2-Xe-134 | 5 | 3 | 3 | 6 | 4 | 2 | 2344.7035 | 2344.7042 | -0.0007 | 0.00073 |
| CO2-Xe-136 | 5 | 3 | 2 | 6 | 4 | 3 | 2344.7035 | 2344.7047 | -0.0012 | 0.00062 |
| CO2-Xe-136 | 5 | 3 | 3 | 6 | 4 | 2 | 2344.7035 | 2344.7047 | -0.0012 | 0.00062 |
| | | | | | | | Blend | 2344.7036 | -0.0001 | |
| CO2-Xe-132 | 4 | 3 | 1 | 5 | 4 | 2 | 2344.7707 | 2344.7709 | -0.0002 | 0.00241 |
| CO2-Xe-132 | 4 | 3 | 2 | 5 | 4 | 1 | 2344.7707 | 2344.7709 | -0.0002 | 0.00241 |
| CO2-Xe-129 | 4 | 3 | 1 | 5 | 4 | 2 | 2344.7707 | 2344.7705 | 0.0002 | 0.00236 |
| CO2-Xe-129 | 4 | 3 | 2 | 5 | 4 | 1 | 2344.7707 | 2344.7705 | 0.0002 | 0.00236 |
| CO2-Xe-131 | 4 | 3 | 1 | 5 | 4 | 2 | 2344.7707 | 2344.7708 | 0.0000 | 0.0019 |
| CO2-Xe-131 | 4 | 3 | 2 | 5 | 4 | 1 | 2344.7707 | 2344.7708 | 0.0000 | 0.0019 |
| CO2-Xe-134 | 4 | 3 | 1 | 5 | 4 | 2 | 2344.7707 | 2344.7711 | -0.0004 | 0.00093 |
| CO2-Xe-134 | 4 | 3 | 2 | 5 | 4 | 1 | 2344.7707 | 2344.7711 | -0.0004 | 0.00093 |
| CO2-Xe-136 | 4 | 3 | 1 | 5 | 4 | 2 | 2344.7707 | 2344.7714 | -0.0006 | 0.00079 |
| CO2-Xe-136 | 4 | 3 | 2 | 5 | 4 | 1 | 2344.7707 | 2344.7714 | -0.0006 | 0.00079 |
| | | | | | | | Blend | 2344.7708 | -0.0001 | |
| CO2-Xe-132 | 3 | 3 | 0 | 4 | 4 | 1 | 2344.8379 | 2344.8381 | -0.0002 | 0.00299 |
| CO2-Xe-132 | 3 | 3 | 1 | 4 | 4 | 0 | 2344.8379 | 2344.8381 | -0.0002 | 0.00299 |
| CO2-Xe-129 | 3 | 3 | 0 | 4 | 4 | 1 | 2344.8379 | 2344.8381 | -0.0002 | 0.00293 |
| CO2-Xe-129 | 3 | 3 | 1 | 4 | 4 | 0 | 2344.8379 | 2344.8381 | -0.0002 | 0.00293 |
| CO2-Xe-131 | 3 | 3 | 0 | 4 | 4 | 1 | 2344.8379 | 2344.8381 | -0.0002 | 0.00235 |
| CO2-Xe-131 | 3 | 3 | 1 | 4 | 4 | 0 | 2344.8379 | 2344.8381 | -0.0002 | 0.00235 |
| CO2-Xe-134 | 3 | 3 | 0 | 4 | 4 | 1 | 2344.8379 | 2344.8381 | -0.0002 | 0.00116 |
| CO2-Xe-134 | 3 | 3 | 1 | 4 | 4 | 0 | 2344.8379 | 2344.8381 | -0.0002 | 0.00116 |
| CO2-Xe-136 | 3 | 3 | 0 | 4 | 4 | 1 | 2344.8379 | 2344.8381 | -0.0002 | 0.00098 |
| CO2-Xe-136 | 3 | 3 | 1 | 4 | 4 | 0 | 2344.8379 | 2344.8381 | -0.0002 | 0.00098 |
| | | | | | | | Blend | 2344.8381 | -0.0002 | |
| CO2-Xe-132 | 6 | 3 | 3 | 6 | 4 | 2 | 2345.1093 | 2345.1086 | 0.0007 | 0.00117 |
| CO2-Xe-132 | 6 | 3 | 4 | 6 | 4 | 3 | 2345.1093 | 2345.1086 | 0.0007 | 0.00117 |
| CO2-Xe-129 | 6 | 3 | 3 | 6 | 4 | 2 | 2345.1093 | 2345.1101 | -0.0008 | 0.00114 |
| CO2-Xe-129 | 6 | 3 | 4 | 6 | 4 | 3 | 2345.1093 | 2345.1101 | -0.0008 | 0.00114 |
| CO2-Xe-132 | 5 | 3 | 2 | 5 | 4 | 1 | 2345.1093 | 2345.1083 | 0.0010 | 0.00111 |
| CO2-Xe-132 | 5 | 3 | 3 | 5 | 4 | 2 | 2345.1093 | 2345.1083 | 0.0010 | 0.00111 |
| CO2-Xe-129 | 5 | 3 | 2 | 5 | 4 | 1 | 2345.1093 | 2345.1098 | -0.0005 | 0.00109 |
| CO2-Xe-129 | 5 | 3 | 3 | 5 | 4 | 2 | 2345.1093 | 2345.1098 | -0.0005 | 0.00109 |
| CO2-Xe-132 | 7 | 3 | 4 | 7 | 4 | 3 | 2345.1093 | 2345.1090 | 0.0004 | 0.00106 |

| | | | | | | | | | |
|---|---|---|---|---|---|---|---|---|---|
| CO2-Xe-132 | 7 | 3 | 5 | 7 | 4 | 4 | 2345.1093 | 2345.1090 | 0.0004 | 0.00106 |
| | | | | | | | Blend | 2345.1091 | 0.0002 | |
| CO2-Xe-132 | 6 | 1 | 6 | 7 | 2 | 5 | 2346.0739 | 2346.0751 | -0.0012 | 0.01925 |
| CO2-Xe-129 | 6 | 1 | 6 | 7 | 2 | 5 | 2346.0739 | 2346.0729 | 0.0010 | 0.01874 |
| CO2-Xe-131 | 6 | 1 | 6 | 7 | 2 | 5 | 2346.0739 | 2346.0744 | -0.0005 | 0.01513 |
| CO2-Xe-134 | 6 | 1 | 6 | 7 | 2 | 5 | 2346.0739 | 2346.0766 | -0.0027 | 0.00748 |
| | | | | | | | Blend | 2346.0744 | -0.0005 | |
| CO2-Xe-132 | 7 | 1 | 6 | 8 | 2 | 7 | 2346.0842 | 2346.0849 | -0.0007 | 0.01575 |
| CO2-Xe-129 | 7 | 1 | 6 | 8 | 2 | 7 | 2346.0842 | 2346.0831 | 0.0011 | 0.01532 |
| CO2-Xe-131 | 7 | 1 | 6 | 8 | 2 | 7 | 2346.0842 | 2346.0843 | -0.0001 | 0.01237 |
| CO2-Xe-134 | 7 | 1 | 6 | 8 | 2 | 7 | 2346.0842 | 2346.0860 | -0.0018 | 0.00612 |
| | | | | | | | Blend | 2346.0843 | -0.0001 | |
| CO2-Xe-132 | 6 | 1 | 5 | 7 | 2 | 6 | 2346.1410 | 2346.1415 | -0.0005 | 0.02111 |
| CO2-Xe-129 | 6 | 1 | 5 | 7 | 2 | 6 | 2346.1410 | 2346.1400 | 0.0010 | 0.02058 |
| CO2-Xe-131 | 6 | 1 | 5 | 7 | 2 | 6 | 2346.1410 | 2346.1410 | 0.0000 | 0.0166 |
| CO2-Xe-134 | 6 | 1 | 5 | 7 | 2 | 6 | 2346.1410 | 2346.1425 | -0.0015 | 0.0082 |
| CO2-Xe-136 | 6 | 1 | 5 | 7 | 2 | 6 | 2346.1410 | 2346.1434 | -0.0025 | 0.00697 |
| | | | | | | | Blend | 2346.1413 | -0.0003 | |
| CO2-Xe-132 | 5 | 1 | 5 | 6 | 2 | 4 | 2346.1521 | 2346.1527 | -0.0006 | 0.0251 |
| CO2-Xe-129 | 5 | 1 | 5 | 6 | 2 | 4 | 2346.1521 | 2346.1509 | 0.0012 | 0.02449 |
| CO2-Xe-131 | 5 | 1 | 5 | 6 | 2 | 4 | 2346.1521 | 2346.1521 | 0.0000 | 0.01974 |
| CO2-Xe-134 | 5 | 1 | 5 | 6 | 2 | 4 | 2346.1521 | 2346.1538 | -0.0017 | 0.00974 |
| | | | | | | | Blend | 2346.1521 | 0.0000 | |
| CO2-Xe-132 | 5 | 1 | 4 | 6 | 2 | 5 | 2346.1994 | 2346.1997 | -0.0003 | 0.02679 |
| CO2-Xe-129 | 5 | 1 | 4 | 6 | 2 | 5 | 2346.1994 | 2346.1985 | 0.0009 | 0.02616 |
| CO2-Xe-131 | 5 | 1 | 4 | 6 | 2 | 5 | 2346.1994 | 2346.1993 | 0.0001 | 0.02108 |
| CO2-Xe-134 | 5 | 1 | 4 | 6 | 2 | 5 | 2346.1994 | 2346.2005 | -0.0011 | 0.01039 |
| CO2-Xe-136 | 5 | 1 | 4 | 6 | 2 | 5 | 2346.1994 | 2346.2013 | -0.0019 | 0.00882 |
| | | | | | | | Blend | 2346.1995 | -0.0001 | |
| CO2-Xe-132 | 4 | 1 | 4 | 5 | 2 | 3 | 2346.2280 | 2346.2283 | -0.0003 | 0.03081 |
| CO2-Xe-129 | 4 | 1 | 4 | 5 | 2 | 3 | 2346.2280 | 2346.2270 | 0.0010 | 0.03011 |
| CO2-Xe-131 | 4 | 1 | 4 | 5 | 2 | 3 | 2346.2280 | 2346.2279 | 0.0001 | 0.02425 |
| CO2-Xe-134 | 4 | 1 | 4 | 5 | 2 | 3 | 2346.2280 | 2346.2291 | -0.0012 | 0.01194 |
| CO2-Xe-136 | 4 | 1 | 4 | 5 | 2 | 3 | 2346.2280 | 2346.2299 | -0.0020 | 0.01013 |
| | | | | | | | Blend | 2346.2281 | -0.0001 | |
| CO2-Xe-132 | 4 | 1 | 3 | 5 | 2 | 4 | 2346.2592 | 2346.2595 | -0.0003 | 0.03216 |
| CO2-Xe-129 | 4 | 1 | 3 | 5 | 2 | 4 | 2346.2592 | 2346.2585 | 0.0007 | 0.03145 |
| CO2-Xe-131 | 4 | 1 | 3 | 5 | 2 | 4 | 2346.2592 | 2346.2591 | 0.0000 | 0.02532 |
| CO2-Xe-134 | 4 | 1 | 3 | 5 | 2 | 4 | 2346.2592 | 2346.2601 | -0.0009 | 0.01246 |
| CO2-Xe-136 | 4 | 1 | 3 | 5 | 2 | 4 | 2346.2592 | 2346.2607 | -0.0015 | 0.01057 |
| | | | | | | | Blend | 2346.2593 | -0.0001 | |
| CO2-Xe-132 | 3 | 1 | 3 | 4 | 2 | 2 | 2346.3021 | 2346.3022 | -0.0001 | 0.0356 |
| CO2-Xe-129 | 3 | 1 | 3 | 4 | 2 | 2 | 2346.3021 | 2346.3013 | 0.0007 | 0.03486 |
| CO2-Xe-131 | 3 | 1 | 3 | 4 | 2 | 2 | 2346.3021 | 2346.3019 | 0.0001 | 0.02804 |
| CO2-Xe-134 | 3 | 1 | 3 | 4 | 2 | 2 | 2346.3021 | 2346.3027 | -0.0007 | 0.01379 |
| CO2-Xe-136 | 3 | 1 | 3 | 4 | 2 | 2 | 2346.3021 | 2346.3033 | -0.0012 | 0.01168 |
| | | | | | | | Blend | 2346.3020 | 0.0000 | |
| CO2-Xe-132 | 3 | 1 | 2 | 4 | 2 | 3 | 2346.3207 | 2346.3208 | -0.0001 | 0.03652 |

| | | | | | | | | | |
|---|---|---|---|---|---|---|---|---|---|
| CO2-Xe-129 | 3 | 1 | 2 | 4 | 2 | 3 | 2346.3207 | 2346.3201 | 0.0006 | 0.03577 |
| CO2-Xe-131 | 3 | 1 | 2 | 4 | 2 | 3 | 2346.3207 | 2346.3206 | 0.0001 | 0.02877 |
| CO2-Xe-134 | 3 | 1 | 2 | 4 | 2 | 3 | 2346.3207 | 2346.3212 | -0.0005 | 0.01414 |
| CO2-Xe-136 | 3 | 1 | 2 | 4 | 2 | 3 | 2346.3207 | 2346.3216 | -0.0009 | 0.01198 |
| | | | | | Blend | | 2346.3207 | 0.0000 | |
| CO2-Xe-132 | 2 | 1 | 2 | 3 | 2 | 1 | 2346.3744 | 2346.3744 | 0.0000 | 0.03889 |
| CO2-Xe-129 | 2 | 1 | 2 | 3 | 2 | 1 | 2346.3744 | 2346.3740 | 0.0004 | 0.03812 |
| CO2-Xe-131 | 2 | 1 | 2 | 3 | 2 | 1 | 2346.3744 | 2346.3742 | 0.0001 | 0.03064 |
| CO2-Xe-134 | 2 | 1 | 2 | 3 | 2 | 1 | 2346.3744 | 2346.3746 | -0.0003 | 0.01505 |
| CO2-Xe-136 | 2 | 1 | 2 | 3 | 2 | 1 | 2346.3744 | 2346.3749 | -0.0005 | 0.01274 |
| | | | | | Blend | | 2346.3743 | 0.0001 | |
| CO2-Xe-132 | 2 | 1 | 1 | 3 | 2 | 2 | 2346.3836 | 2346.3836 | 0.0000 | 0.03938 |
| CO2-Xe-129 | 2 | 1 | 1 | 3 | 2 | 2 | 2346.3836 | 2346.3833 | 0.0003 | 0.03861 |
| CO2-Xe-131 | 2 | 1 | 1 | 3 | 2 | 2 | 2346.3836 | 2346.3835 | 0.0001 | 0.03103 |
| CO2-Xe-134 | 2 | 1 | 1 | 3 | 2 | 2 | 2346.3836 | 2346.3838 | -0.0002 | 0.01524 |
| CO2-Xe-136 | 2 | 1 | 1 | 3 | 2 | 2 | 2346.3836 | 2346.3840 | -0.0004 | 0.0129 |
| | | | | | Blend | | 2346.3836 | 0.0000 | |
| CO2-Xe-132 | 1 | 1 | 1 | 2 | 2 | 0 | 2346.4450 | 2346.4450 | 0.0000 | 0.04083 |
| CO2-Xe-129 | 1 | 1 | 1 | 2 | 2 | 0 | 2346.4450 | 2346.4449 | 0.0000 | 0.04006 |
| CO2-Xe-131 | 1 | 1 | 1 | 2 | 2 | 0 | 2346.4450 | 2346.4450 | 0.0000 | 0.03218 |
| CO2-Xe-134 | 1 | 1 | 1 | 2 | 2 | 0 | 2346.4450 | 2346.4450 | 0.0000 | 0.01579 |
| CO2-Xe-136 | 1 | 1 | 1 | 2 | 2 | 0 | 2346.4450 | 2346.4450 | 0.0000 | 0.01336 |
| | | | | | Blend | | 2346.4450 | 0.0000 | |
| CO2-Xe-132 | 1 | 1 | 0 | 2 | 2 | 1 | 2346.4480 | 2346.4480 | 0.0000 | 0.041 |
| CO2-Xe-129 | 1 | 1 | 0 | 2 | 2 | 1 | 2346.4480 | 2346.4480 | 0.0000 | 0.04023 |
| CO2-Xe-131 | 1 | 1 | 0 | 2 | 2 | 1 | 2346.4480 | 2346.4480 | 0.0000 | 0.03231 |
| CO2-Xe-134 | 1 | 1 | 0 | 2 | 2 | 1 | 2346.4480 | 2346.4480 | 0.0000 | 0.01586 |
| CO2-Xe-136 | 1 | 1 | 0 | 2 | 2 | 1 | 2346.4480 | 2346.4480 | 0.0000 | 0.01342 |
| | | | | | Blend | | 2346.4480 | 0.0000 | |
| CO2-Xe-132 | 11 | 1 | 11 | 11 | 2 | 10 | 2346.4810 | 2346.4809 | 0.0001 | 0.00682 |
| CO2-Xe-129 | 11 | 1 | 11 | 11 | 2 | 10 | 2346.4810 | 2346.4806 | 0.0004 | 0.00657 |
| CO2-Xe-131 | 11 | 1 | 11 | 11 | 2 | 10 | 2346.4810 | 2346.4808 | 0.0002 | 0.00534 |
| CO2-Xe-134 | 11 | 1 | 11 | 11 | 2 | 10 | 2346.4810 | 2346.4811 | -0.0001 | 0.00267 |
| CO2-Xe-136 | 11 | 1 | 11 | 11 | 2 | 10 | 2346.4810 | 2346.4813 | -0.0003 | 0.00228 |
| | | | | | Blend | | 2346.4809 | 0.0001 | |
| CO2-Xe-132 | 10 | 1 | 10 | 10 | 2 | 9 | 2346.4979 | 2346.4978 | 0.0001 | 0.0107 |
| CO2-Xe-129 | 10 | 1 | 10 | 10 | 2 | 9 | 2346.4979 | 2346.4977 | 0.0002 | 0.01035 |
| CO2-Xe-131 | 10 | 1 | 10 | 10 | 2 | 9 | 2346.4979 | 2346.4978 | 0.0001 | 0.00839 |
| CO2-Xe-134 | 10 | 1 | 10 | 10 | 2 | 9 | 2346.4979 | 2346.4979 | 0.0000 | 0.00418 |
| CO2-Xe-136 | 10 | 1 | 10 | 10 | 2 | 9 | 2346.4979 | 2346.4980 | -0.0001 | 0.00357 |
| | | | | | Blend | | 2346.4978 | 0.0001 | |
| CO2-Xe-132 | 9 | 1 | 9 | 9 | 2 | 8 | 2346.5132 | 2346.5132 | 0.0000 | 0.01582 |
| CO2-Xe-129 | 9 | 1 | 9 | 9 | 2 | 8 | 2346.5132 | 2346.5132 | 0.0000 | 0.01534 |
| CO2-Xe-131 | 9 | 1 | 9 | 9 | 2 | 8 | 2346.5132 | 2346.5132 | 0.0000 | 0.01242 |
| CO2-Xe-134 | 9 | 1 | 9 | 9 | 2 | 8 | 2346.5132 | 2346.5131 | 0.0000 | 0.00616 |
| CO2-Xe-136 | 9 | 1 | 9 | 9 | 2 | 8 | 2346.5132 | 2346.5131 | 0.0001 | 0.00526 |
| | | | | | Blend | | 2346.5132 | 0.0000 | |
| CO2-Xe-132 | 8 | 1 | 8 | 8 | 2 | 7 | 2346.5271 | 2346.5269 | 0.0002 | 0.02199 |

| | | | | | | | | | |
|---|---|---|---|---|---|---|---|---|---|
| CO2-Xe-129 | 8 | 1 | 8 | 8 | 2 | 7 | 2346.5271 | 2346.5271 | 0.0000 | 0.02137 |
| CO2-Xe-131 | 8 | 1 | 8 | 8 | 2 | 7 | 2346.5271 | 2346.5270 | 0.0001 | 0.01727 |
| CO2-Xe-134 | 8 | 1 | 8 | 8 | 2 | 7 | 2346.5271 | 2346.5268 | 0.0003 | 0.00855 |
| CO2-Xe-136 | 8 | 1 | 8 | 8 | 2 | 7 | 2346.5271 | 2346.5267 | 0.0004 | 0.00728 |
| | | | | | Blend | | | 2346.5269 | 0.0001 | |
| CO2-Xe-132 | 7 | 1 | 7 | 7 | 2 | 6 | 2346.5393 | 2346.5391 | 0.0002 | 0.02861 |
| CO2-Xe-129 | 7 | 1 | 7 | 7 | 2 | 6 | 2346.5393 | 2346.5394 | -0.0001 | 0.02787 |
| CO2-Xe-131 | 7 | 1 | 7 | 7 | 2 | 6 | 2346.5393 | 2346.5392 | 0.0001 | 0.02249 |
| CO2-Xe-134 | 7 | 1 | 7 | 7 | 2 | 6 | 2346.5393 | 2346.5389 | 0.0003 | 0.01111 |
| CO2-Xe-136 | 7 | 1 | 7 | 7 | 2 | 6 | 2346.5393 | 2346.5387 | 0.0005 | 0.00944 |
| | | | | | Blend | | | 2346.5391 | 0.0001 | |
| CO2-Xe-132 | 6 | 1 | 6 | 6 | 2 | 5 | 2346.5500 | 2346.5497 | 0.0002 | 0.03465 |
| CO2-Xe-129 | 6 | 1 | 6 | 6 | 2 | 5 | 2346.5500 | 2346.5502 | -0.0002 | 0.03382 |
| CO2-Xe-131 | 6 | 1 | 6 | 6 | 2 | 5 | 2346.5500 | 2346.5499 | 0.0001 | 0.02726 |
| CO2-Xe-134 | 6 | 1 | 6 | 6 | 2 | 5 | 2346.5500 | 2346.5495 | 0.0005 | 0.01344 |
| CO2-Xe-136 | 6 | 1 | 6 | 6 | 2 | 5 | 2346.5500 | 2346.5492 | 0.0007 | 0.01141 |
| | | | | | Blend | | | 2346.5498 | 0.0001 | |
| CO2-Xe-132 | 5 | 1 | 5 | 5 | 2 | 4 | 2346.5590 | 2346.5588 | 0.0001 | 0.03869 |
| CO2-Xe-129 | 5 | 1 | 5 | 5 | 2 | 4 | 2346.5590 | 2346.5593 | -0.0004 | 0.03783 |
| CO2-Xe-131 | 5 | 1 | 5 | 5 | 2 | 4 | 2346.5590 | 2346.5590 | 0.0000 | 0.03045 |
| CO2-Xe-134 | 5 | 1 | 5 | 5 | 2 | 4 | 2346.5590 | 2346.5585 | 0.0005 | 0.01499 |
| CO2-Xe-136 | 5 | 1 | 5 | 5 | 2 | 4 | 2346.5590 | 2346.5582 | 0.0008 | 0.01271 |
| | | | | | Blend | | | 2346.5589 | 0.0001 | |
| CO2-Xe-132 | 4 | 1 | 4 | 4 | 2 | 3 | 2346.5667 | 2346.5664 | 0.0003 | 0.03911 |
| CO2-Xe-129 | 4 | 1 | 4 | 4 | 2 | 3 | 2346.5667 | 2346.5670 | -0.0003 | 0.0383 |
| CO2-Xe-131 | 4 | 1 | 4 | 4 | 2 | 3 | 2346.5667 | 2346.5666 | 0.0001 | 0.0308 |
| CO2-Xe-134 | 4 | 1 | 4 | 4 | 2 | 3 | 2346.5667 | 2346.5660 | 0.0007 | 0.01514 |
| CO2-Xe-136 | 4 | 1 | 4 | 4 | 2 | 3 | 2346.5667 | 2346.5657 | 0.0010 | 0.01283 |
| | | | | | Blend | | | 2346.5665 | 0.0002 | |
| CO2-Xe-132 | 3 | 1 | 3 | 3 | 2 | 2 | 2346.5727 | 2346.5725 | 0.0002 | 0.03435 |
| CO2-Xe-129 | 3 | 1 | 3 | 3 | 2 | 2 | 2346.5727 | 2346.5731 | -0.0004 | 0.03368 |
| CO2-Xe-131 | 3 | 1 | 3 | 3 | 2 | 2 | 2346.5727 | 2346.5727 | 0.0000 | 0.02706 |
| CO2-Xe-134 | 3 | 1 | 3 | 3 | 2 | 2 | 2346.5727 | 2346.5720 | 0.0007 | 0.01329 |
| CO2-Xe-136 | 3 | 1 | 3 | 3 | 2 | 2 | 2346.5727 | 2346.5716 | 0.0011 | 0.01125 |
| | | | | | Blend | | | 2346.5726 | 0.0001 | |
| CO2-Xe-132 | 2 | 1 | 2 | 2 | 2 | 1 | 2346.5772 | 2346.5770 | 0.0002 | 0.02278 |
| CO2-Xe-129 | 2 | 1 | 2 | 2 | 2 | 1 | 2346.5772 | 2346.5777 | -0.0005 | 0.02235 |
| CO2-Xe-131 | 2 | 1 | 2 | 2 | 2 | 1 | 2346.5772 | 2346.5772 | 0.0000 | 0.01795 |
| CO2-Xe-134 | 2 | 1 | 2 | 2 | 2 | 1 | 2346.5772 | 2346.5765 | 0.0007 | 0.00881 |
| CO2-Xe-136 | 2 | 1 | 2 | 2 | 2 | 1 | 2346.5772 | 2346.5761 | 0.0011 | 0.00745 |
| | | | | | Blend | | | 2346.5771 | 0.0001 | |
| CO2-Xe-132 | 2 | 1 | 1 | 2 | 2 | 0 | 2346.5864 | 2346.5861 | 0.0003 | 0.02307 |
| CO2-Xe-129 | 2 | 1 | 1 | 2 | 2 | 0 | 2346.5864 | 2346.5869 | -0.0005 | 0.02263 |
| CO2-Xe-131 | 2 | 1 | 1 | 2 | 2 | 0 | 2346.5864 | 2346.5864 | 0.0000 | 0.01818 |
| CO2-Xe-134 | 2 | 1 | 1 | 2 | 2 | 0 | 2346.5864 | 2346.5856 | 0.0008 | 0.00892 |
| CO2-Xe-136 | 2 | 1 | 1 | 2 | 2 | 0 | 2346.5864 | 2346.5851 | 0.0013 | 0.00755 |
| | | | | | Blend | | | 2346.5862 | 0.0001 | |
| CO2-Xe-132 | 3 | 1 | 2 | 3 | 2 | 1 | 2346.5909 | 2346.5907 | 0.0003 | 0.03521 |

| | | | | | | | | | |
|---|---|---|---|---|---|---|---|---|---|
| CO2-Xe-129 | 3 | 1 | 2 | 3 | 2 | 1 | 2346.5909 | 2346.5915 | -0.0006 | 0.03453 |
| CO2-Xe-131 | 3 | 1 | 2 | 3 | 2 | 1 | 2346.5909 | 2346.5909 | 0.0000 | 0.02774 |
| CO2-Xe-134 | 3 | 1 | 2 | 3 | 2 | 1 | 2346.5909 | 2346.5901 | 0.0008 | 0.01362 |
| CO2-Xe-136 | 3 | 1 | 2 | 3 | 2 | 1 | 2346.5909 | 2346.5896 | 0.0014 | 0.01153 |
| | | | | | | Blend | 2346.5908 | 0.0001 | |
| CO2-Xe-132 | 4 | 1 | 3 | 4 | 2 | 2 | 2346.5970 | 2346.5966 | 0.0004 | 0.04075 |
| CO2-Xe-129 | 4 | 1 | 3 | 4 | 2 | 2 | 2346.5970 | 2346.5975 | -0.0005 | 0.03992 |
| CO2-Xe-131 | 4 | 1 | 3 | 4 | 2 | 2 | 2346.5970 | 2346.5969 | 0.0001 | 0.03209 |
| CO2-Xe-134 | 4 | 1 | 3 | 4 | 2 | 2 | 2346.5970 | 2346.5960 | 0.0010 | 0.01577 |
| CO2-Xe-136 | 4 | 1 | 3 | 4 | 2 | 2 | 2346.5970 | 2346.5955 | 0.0016 | 0.01336 |
| | | | | | | Blend | 2346.5968 | 0.0003 | |
| CO2-Xe-132 | 5 | 1 | 4 | 5 | 2 | 3 | 2346.6041 | 2346.6039 | 0.0002 | 0.04111 |
| CO2-Xe-129 | 5 | 1 | 4 | 5 | 2 | 3 | 2346.6041 | 2346.6049 | -0.0008 | 0.04023 |
| CO2-Xe-131 | 5 | 1 | 4 | 5 | 2 | 3 | 2346.6041 | 2346.6042 | -0.0001 | 0.03237 |
| CO2-Xe-134 | 5 | 1 | 4 | 5 | 2 | 3 | 2346.6041 | 2346.6033 | 0.0008 | 0.01592 |
| CO2-Xe-136 | 5 | 1 | 4 | 5 | 2 | 3 | 2346.6041 | 2346.6027 | 0.0015 | 0.0135 |
| | | | | | | Blend | 2346.6041 | 0.0001 | |
| CO2-Xe-132 | 6 | 1 | 5 | 6 | 2 | 4 | 2346.6128 | 2346.6124 | 0.0003 | 0.03769 |
| CO2-Xe-129 | 6 | 1 | 5 | 6 | 2 | 4 | 2346.6128 | 2346.6135 | -0.0007 | 0.03682 |
| CO2-Xe-131 | 6 | 1 | 5 | 6 | 2 | 4 | 2346.6128 | 2346.6128 | 0.0000 | 0.02966 |
| CO2-Xe-134 | 6 | 1 | 5 | 6 | 2 | 4 | 2346.6128 | 2346.6117 | 0.0010 | 0.01461 |
| CO2-Xe-136 | 6 | 1 | 5 | 6 | 2 | 4 | 2346.6128 | 2346.6111 | 0.0017 | 0.0124 |
| | | | | | | Blend | 2346.6126 | 0.0002 | |
| CO2-Xe-132 | 7 | 1 | 6 | 7 | 2 | 5 | 2346.6225 | 2346.6221 | 0.0004 | 0.03195 |
| CO2-Xe-129 | 7 | 1 | 6 | 7 | 2 | 5 | 2346.6225 | 2346.6232 | -0.0007 | 0.03117 |
| CO2-Xe-131 | 7 | 1 | 6 | 7 | 2 | 5 | 2346.6225 | 2346.6224 | 0.0001 | 0.02513 |
| CO2-Xe-134 | 7 | 1 | 6 | 7 | 2 | 5 | 2346.6225 | 2346.6213 | 0.0012 | 0.0124 |
| CO2-Xe-136 | 7 | 1 | 6 | 7 | 2 | 5 | 2346.6225 | 2346.6206 | 0.0019 | 0.01053 |
| | | | | | | Blend | 2346.6223 | 0.0003 | |
| CO2-Xe-132 | 8 | 1 | 7 | 8 | 2 | 6 | 2346.6330 | 2346.6327 | 0.0004 | 0.02529 |
| CO2-Xe-129 | 8 | 1 | 7 | 8 | 2 | 6 | 2346.6330 | 2346.6339 | -0.0009 | 0.02462 |
| CO2-Xe-131 | 8 | 1 | 7 | 8 | 2 | 6 | 2346.6330 | 2346.6331 | 0.0000 | 0.01988 |
| CO2-Xe-134 | 8 | 1 | 7 | 8 | 2 | 6 | 2346.6330 | 2346.6319 | 0.0012 | 0.00983 |
| | | | | | | Blend | 2346.6330 | 0.0000 | |
| CO2-Xe-132 | 9 | 1 | 8 | 9 | 2 | 7 | 2346.6445 | 2346.6441 | 0.0004 | 0.01879 |
| CO2-Xe-129 | 9 | 1 | 8 | 9 | 2 | 7 | 2346.6445 | 2346.6454 | -0.0009 | 0.01825 |
| CO2-Xe-131 | 9 | 1 | 8 | 9 | 2 | 7 | 2346.6445 | 2346.6445 | 0.0000 | 0.01476 |
| CO2-Xe-134 | 9 | 1 | 8 | 9 | 2 | 7 | 2346.6445 | 2346.6432 | 0.0013 | 0.00731 |
| | | | | | | Blend | 2346.6445 | 0.0000 | |
| CO2-Xe-132 | 10 | 1 | 9 | 10 | 2 | 8 | 2346.6570 | 2346.6561 | 0.0009 | 0.01314 |
| CO2-Xe-129 | 10 | 1 | 9 | 10 | 2 | 8 | 2346.6570 | 2346.6575 | -0.0005 | 0.01273 |
| CO2-Xe-131 | 10 | 1 | 9 | 10 | 2 | 8 | 2346.6570 | 2346.6565 | 0.0004 | 0.01031 |
| CO2-Xe-134 | 10 | 1 | 9 | 10 | 2 | 8 | 2346.6570 | 2346.6552 | 0.0018 | 0.00512 |
| | | | | | | Blend | 2346.6565 | 0.0004 | |
| CO2-Xe-132 | 11 | 1 | 10 | 11 | 2 | 9 | 2346.6688 | 2346.6685 | 0.0003 | 0.00868 |
| CO2-Xe-129 | 11 | 1 | 10 | 11 | 2 | 9 | 2346.6688 | 2346.6700 | -0.0012 | 0.00838 |
| CO2-Xe-131 | 11 | 1 | 10 | 11 | 2 | 9 | 2346.6688 | 2346.6690 | -0.0002 | 0.0068 |
| CO2-Xe-134 | 11 | 1 | 10 | 11 | 2 | 9 | 2346.6688 | 2346.6675 | 0.0013 | 0.00339 |

| | | | | | | | Blend | 2346.6690 | −0.0001 | |
|---|---|---|---|---|---|---|---|---|---|---|
| CO2-Xe-132 | 10 | 1 | 10 | 9 | 2 | 7 | 2347.1666 | 2347.1657 | 0.0009 | 0.00545 |
| CO2-Xe-131 | 10 | 1 | 10 | 9 | 2 | 7 | 2347.1666 | 2347.1668 | −0.0002 | 0.00427 |
| | | | | | | | Blend | 2347.1662 | 0.0004 | |
| CO2-Xe-129 | 10 | 1 | 10 | 9 | 2 | 7 | 2347.1696 | 2347.1691 | 0.0005 | 0.00527 |
| CO2-Xe-136 | 8 | 1 | 7 | 7 | 2 | 6 | 2347.1696 | 2347.1707 | −0.0011 | 0.00373 |
| | | | | | | | Blend | 2347.1698 | −0.0002 | |
| CO2-Xe-132 | 8 | 1 | 7 | 7 | 2 | 6 | 2347.1769 | 2347.1762 | 0.0007 | 0.0113 |
| CO2-Xe-131 | 8 | 1 | 7 | 7 | 2 | 6 | 2347.1769 | 2347.1776 | −0.0007 | 0.00889 |
| | | | | | | | Blend | 2347.1768 | 0.0001 | |
| CO2-Xe-129 | 8 | 1 | 7 | 7 | 2 | 6 | 2347.1800 | 2347.1805 | −0.0005 | |
| CO2-Xe-129 | 10 | 1 | 10 | 11 | 0 | 11 | 2347.2106 | 2347.2104 | 0.0002 | |
| CO2-Xe-132 | 10 | 1 | 10 | 11 | 0 | 11 | 2347.2145 | 2347.2151 | −0.0006 | 0.0213 |
| CO2-Xe-131 | 10 | 1 | 10 | 11 | 0 | 11 | 2347.2145 | 2347.2136 | 0.0009 | 0.0167 |
| | | | | | | | Blend | 2347.2144 | 0.0000 | |
| CO2-Xe-136 | 10 | 1 | 10 | 11 | 0 | 11 | 2347.2214 | 2347.2211 | 0.0002 | |
| CO2-Xe-134 | 9 | 1 | 8 | 8 | 2 | 7 | 2347.2541 | 2347.2540 | 0.0001 | 0.00357 |
| CO2-Xe-134 | 12 | 1 | 12 | 11 | 2 | 9 | 2347.2541 | 2347.2550 | −0.0009 | 0.0009 |
| | | | | | | | Blend | 2347.2542 | −0.0001 | |
| CO2-Xe-132 | 9 | 1 | 8 | 8 | 2 | 7 | 2347.2580 | 2347.2571 | 0.0009 | 0.00919 |
| CO2-Xe-131 | 9 | 1 | 8 | 8 | 2 | 7 | 2347.2580 | 2347.2587 | −0.0007 | 0.00723 |
| CO2-Xe-132 | 12 | 1 | 12 | 11 | 2 | 9 | 2347.2580 | 2347.2574 | 0.0007 | 0.0023 |
| | | | | | | | Blend | 2347.2578 | 0.0003 | |
| CO2-Xe-129 | 9 | 1 | 8 | 8 | 2 | 7 | 2347.2616 | 2347.2619 | −0.0003 | |
| CO2-Xe-129 | 9 | 1 | 9 | 10 | 0 | 10 | 2347.2902 | 2347.2901 | 0.0001 | |
| CO2-Xe-132 | 9 | 1 | 9 | 10 | 0 | 10 | 2347.2937 | 2347.2944 | −0.0007 | 0.03193 |
| CO2-Xe-131 | 9 | 1 | 9 | 10 | 0 | 10 | 2347.2937 | 2347.2930 | 0.0007 | 0.02505 |
| | | | | | | | Blend | 2347.2938 | −0.0001 | |
| CO2-Xe-136 | 9 | 1 | 9 | 10 | 0 | 10 | 2347.2996 | 2347.2998 | −0.0002 | |
| CO2-Xe-129 | 8 | 1 | 8 | 9 | 0 | 9 | 2347.3693 | 2347.3692 | 0.0001 | |
| CO2-Xe-131 | 8 | 1 | 8 | 9 | 0 | 9 | 2347.3723 | 2347.3718 | 0.0005 | 0.03543 |
| CO2-Xe-132 | 8 | 1 | 8 | 9 | 0 | 9 | 2347.3723 | 2347.3730 | −0.0007 | 0.04511 |
| | | | | | | | Blend | 2347.3725 | −0.0002 | |
| CO2-Xe-129 | 7 | 1 | 7 | 8 | 0 | 8 | 2347.4477 | 2347.4475 | 0.0003 | |
| CO2-Xe-131 | 7 | 1 | 7 | 8 | 0 | 8 | 2347.4502 | 2347.4497 | 0.0005 | 0.04707 |

| | | | | | | | | | |
|---|---|---|---|---|---|---|---|---|---|
| CO2-Xe-132 | 7 | 1 | 7 | 8 | 0 | 8 | 2347.4502 | 2347.4508 | -0.0006 | 0.05989 |
| | | | Blend | | | | | 2347.4503 | -0.0001 | |
| CO2-Xe-132 | 6 | 1 | 6 | 7 | 0 | 7 | 2347.5270 | 2347.5276 | -0.0007 | 0.07434 |
| CO2-Xe-131 | 6 | 1 | 6 | 7 | 0 | 7 | 2347.5270 | 2347.5267 | 0.0003 | 0.05846 |
| | | | Blend | | | | | 2347.5272 | -0.0002 | |
| CO2-Xe-132 | 5 | 1 | 5 | 6 | 0 | 6 | 2347.6024 | 2347.6034 | -0.0010 | 0.08563 |
| CO2-Xe-129 | 5 | 1 | 5 | 6 | 0 | 6 | 2347.6024 | 2347.6009 | 0.0015 | 0.08361 |
| CO2-Xe-131 | 5 | 1 | 5 | 6 | 0 | 6 | 2347.6024 | 2347.6026 | -0.0002 | 0.06737 |
| | | | Blend | | | | | 2347.6023 | 0.0001 | |
| CO2-Xe-132 | 4 | 1 | 4 | 5 | 0 | 5 | 2347.6772 | 2347.6780 | -0.0008 | 0.09037 |
| CO2-Xe-129 | 4 | 1 | 4 | 5 | 0 | 5 | 2347.6772 | 2347.6759 | 0.0012 | 0.08838 |
| CO2-Xe-131 | 4 | 1 | 4 | 5 | 0 | 5 | 2347.6772 | 2347.6773 | -0.0001 | 0.07114 |
| CO2-Xe-134 | 4 | 1 | 4 | 5 | 0 | 5 | 2347.6772 | 2347.6792 | -0.0021 | 0.03502 |
| | | | Blend | | | | | 2347.6773 | -0.0001 | |
| CO2-Xe-132 | 3 | 1 | 3 | 4 | 0 | 4 | 2347.7506 | 2347.7512 | -0.0007 | 0.08535 |
| CO2-Xe-129 | 3 | 1 | 3 | 4 | 0 | 4 | 2347.7506 | 2347.7496 | 0.0009 | 0.08357 |
| CO2-Xe-131 | 3 | 1 | 3 | 4 | 0 | 4 | 2347.7506 | 2347.7507 | -0.0002 | 0.06722 |
| CO2-Xe-134 | 3 | 1 | 3 | 4 | 0 | 4 | 2347.7506 | 2347.7522 | -0.0017 | 0.03305 |
| CO2-Xe-136 | 3 | 1 | 3 | 4 | 0 | 4 | 2347.7506 | 2347.7532 | -0.0027 | 0.02801 |
| | | | Blend | | | | | 2347.7510 | -0.0004 | |
| CO2-Xe-132 | 2 | 1 | 2 | 3 | 0 | 3 | 2347.8228 | 2347.8232 | -0.0003 | 0.0684 |
| CO2-Xe-129 | 2 | 1 | 2 | 3 | 0 | 3 | 2347.8228 | 2347.8220 | 0.0009 | 0.06703 |
| CO2-Xe-131 | 2 | 1 | 2 | 3 | 0 | 3 | 2347.8228 | 2347.8228 | 0.0001 | 0.05388 |
| CO2-Xe-134 | 2 | 1 | 2 | 3 | 0 | 3 | 2347.8228 | 2347.8239 | -0.0011 | 0.02647 |
| CO2-Xe-136 | 2 | 1 | 2 | 3 | 0 | 3 | 2347.8228 | 2347.8246 | -0.0018 | 0.02242 |
| | | | Blend | | | | | 2347.8230 | -0.0001 | |
| CO2-Xe-132 | 1 | 1 | 1 | 2 | 0 | 2 | 2347.8934 | 2347.8937 | -0.0002 | 0.03924 |
| CO2-Xe-129 | 1 | 1 | 1 | 2 | 0 | 2 | 2347.8934 | 2347.8929 | 0.0005 | 0.03848 |
| CO2-Xe-131 | 1 | 1 | 1 | 2 | 0 | 2 | 2347.8934 | 2347.8934 | 0.0000 | 0.03092 |
| CO2-Xe-134 | 1 | 1 | 1 | 2 | 0 | 2 | 2347.8934 | 2347.8942 | -0.0007 | 0.01518 |
| CO2-Xe-136 | 1 | 1 | 1 | 2 | 0 | 2 | 2347.8934 | 2347.8946 | -0.0012 | 0.01285 |
| | | | Blend | | | | | 2347.8935 | -0.0001 | |
| CO2-Xe-132 | 1 | 1 | 0 | 1 | 0 | 1 | 2348.0319 | 2348.0318 | 0.0001 | 0.12812 |
| CO2-Xe-129 | 1 | 1 | 0 | 1 | 0 | 1 | 2348.0319 | 2348.0318 | 0.0001 | 0.12571 |
| CO2-Xe-131 | 1 | 1 | 0 | 1 | 0 | 1 | 2348.0319 | 2348.0318 | 0.0001 | 0.10096 |
| CO2-Xe-134 | 1 | 1 | 0 | 1 | 0 | 1 | 2348.0319 | 2348.0317 | 0.0001 | 0.04954 |
| CO2-Xe-136 | 1 | 1 | 0 | 1 | 0 | 1 | 2348.0319 | 2348.0317 | 0.0001 | 0.04193 |
| | | | Blend | | | | | 2348.0318 | 0.0001 | |
| CO2-Xe-132 | 2 | 1 | 1 | 2 | 0 | 2 | 2348.0348 | 2348.0348 | 0.0000 | 0.19296 |
| CO2-Xe-129 | 2 | 1 | 1 | 2 | 0 | 2 | 2348.0348 | 2348.0349 | -0.0001 | 0.18921 |
| CO2-Xe-131 | 2 | 1 | 1 | 2 | 0 | 2 | 2348.0348 | 2348.0348 | -0.0001 | 0.15203 |
| CO2-Xe-134 | 2 | 1 | 1 | 2 | 0 | 2 | 2348.0348 | 2348.0348 | 0.0000 | 0.07464 |
| CO2-Xe-136 | 2 | 1 | 1 | 2 | 0 | 2 | 2348.0348 | 2348.0347 | 0.0000 | 0.06319 |
| | | | Blend | | | | | 2348.0348 | -0.0001 | |
| CO2-Xe-132 | 3 | 1 | 2 | 3 | 0 | 3 | 2348.0395 | 2348.0394 | 0.0000 | 0.23205 |
| CO2-Xe-129 | 3 | 1 | 2 | 3 | 0 | 3 | 2348.0395 | 2348.0395 | -0.0001 | 0.22734 |
| CO2-Xe-131 | 3 | 1 | 2 | 3 | 0 | 3 | 2348.0395 | 2348.0395 | 0.0000 | 0.18278 |
| CO2-Xe-134 | 3 | 1 | 2 | 3 | 0 | 3 | 2348.0395 | 2348.0394 | 0.0001 | 0.08981 |

| | | | | | | | | | |
|---|---|---|---|---|---|---|---|---|---|
| CO2-Xe-136 | 3 | 1 | 2 | 3 | 0 | 3 | 2348.0395 | 2348.0393 | 0.0002 | 0.07608 |
| | | | | Blend | | | | 2348.0394 | 0.0000 |
| CO2-Xe-132 | 4 | 1 | 3 | 4 | 0 | 4 | 2348.0457 | 2348.0457 | 0.0001 | 0.24362 |
| CO2-Xe-129 | 4 | 1 | 3 | 4 | 0 | 4 | 2348.0457 | 2348.0458 | -0.0001 | 0.2384 |
| CO2-Xe-131 | 4 | 1 | 3 | 4 | 0 | 4 | 2348.0457 | 2348.0457 | 0.0000 | 0.19182 |
| CO2-Xe-134 | 4 | 1 | 3 | 4 | 0 | 4 | 2348.0457 | 2348.0455 | 0.0002 | 0.09436 |
| CO2-Xe-136 | 4 | 1 | 3 | 4 | 0 | 4 | 2348.0457 | 2348.0454 | 0.0003 | 0.07999 |
| | | | | Blend | | | | 2348.0457 | 0.0000 |
| CO2-Xe-132 | 5 | 1 | 4 | 5 | 0 | 5 | 2348.0536 | 2348.0535 | 0.0001 | 0.23113 |
| CO2-Xe-129 | 5 | 1 | 4 | 5 | 0 | 5 | 2348.0536 | 2348.0538 | -0.0002 | 0.22584 |
| CO2-Xe-131 | 5 | 1 | 4 | 5 | 0 | 5 | 2348.0536 | 2348.0536 | 0.0000 | 0.18189 |
| CO2-Xe-134 | 5 | 1 | 4 | 5 | 0 | 5 | 2348.0536 | 2348.0534 | 0.0002 | 0.08961 |
| CO2-Xe-136 | 5 | 1 | 4 | 5 | 0 | 5 | 2348.0536 | 2348.0532 | 0.0004 | 0.07603 |
| | | | | Blend | | | | 2348.0536 | 0.0000 |
| CO2-Xe-132 | 6 | 1 | 5 | 6 | 0 | 6 | 2348.0632 | 2348.0632 | 0.0001 | 0.20156 |
| CO2-Xe-129 | 6 | 1 | 5 | 6 | 0 | 6 | 2348.0632 | 2348.0635 | -0.0003 | 0.19661 |
| CO2-Xe-131 | 6 | 1 | 5 | 6 | 0 | 6 | 2348.0632 | 2348.0633 | 0.0000 | 0.15853 |
| CO2-Xe-134 | 6 | 1 | 5 | 6 | 0 | 6 | 2348.0632 | 2348.0629 | 0.0003 | 0.07823 |
| CO2-Xe-136 | 6 | 1 | 5 | 6 | 0 | 6 | 2348.0632 | 2348.0627 | 0.0005 | 0.06645 |
| | | | | Blend | | | | 2348.0632 | 0.0000 |
| CO2-Xe-132 | 7 | 1 | 6 | 7 | 0 | 7 | 2348.0747 | 2348.0746 | 0.0001 | 0.16314 |
| CO2-Xe-129 | 7 | 1 | 6 | 7 | 0 | 7 | 2348.0747 | 2348.0751 | -0.0004 | 0.15881 |
| CO2-Xe-131 | 7 | 1 | 6 | 7 | 0 | 7 | 2348.0747 | 2348.0747 | -0.0001 | 0.12823 |
| CO2-Xe-134 | 7 | 1 | 6 | 7 | 0 | 7 | 2348.0747 | 2348.0743 | 0.0004 | 0.06341 |
| CO2-Xe-136 | 7 | 1 | 6 | 7 | 0 | 7 | 2348.0747 | 2348.0739 | 0.0007 | 0.05393 |
| | | | | Blend | | | | 2348.0747 | 0.0000 |
| CO2-Xe-132 | 8 | 1 | 7 | 8 | 0 | 8 | 2348.0880 | 2348.0879 | 0.0001 | 0.1233 |
| CO2-Xe-129 | 8 | 1 | 7 | 8 | 0 | 8 | 2348.0880 | 2348.0885 | -0.0005 | 0.11975 |
| CO2-Xe-131 | 8 | 1 | 7 | 8 | 0 | 8 | 2348.0880 | 2348.0881 | -0.0001 | 0.09684 |
| CO2-Xe-134 | 8 | 1 | 7 | 8 | 0 | 8 | 2348.0880 | 2348.0875 | 0.0006 | 0.04799 |
| CO2-Xe-136 | 8 | 1 | 7 | 8 | 0 | 8 | 2348.0880 | 2348.0871 | 0.0010 | 0.04088 |
| | | | | Blend | | | | 2348.0880 | 0.0000 |
| CO2-Xe-132 | 1 | 1 | 1 | 0 | 0 | 0 | 2348.0963 | 2348.0962 | 0.0001 | 0.08967 |
| CO2-Xe-129 | 1 | 1 | 1 | 0 | 0 | 0 | 2348.0963 | 2348.0966 | -0.0003 | 0.088 |
| CO2-Xe-131 | 1 | 1 | 1 | 0 | 0 | 0 | 2348.0963 | 2348.0964 | 0.0000 | 0.07067 |
| CO2-Xe-134 | 1 | 1 | 1 | 0 | 0 | 0 | 2348.0963 | 2348.0960 | 0.0003 | 0.03467 |
| CO2-Xe-136 | 1 | 1 | 1 | 0 | 0 | 0 | 2348.0963 | 2348.0958 | 0.0006 | 0.02933 |
| | | | | Blend | | | | 2348.0963 | 0.0000 |
| CO2-Xe-132 | 9 | 1 | 8 | 9 | 0 | 9 | 2348.1034 | 2348.1032 | 0.0002 | 0.08738 |
| CO2-Xe-129 | 9 | 1 | 8 | 9 | 0 | 9 | 2348.1034 | 2348.1041 | -0.0007 | 0.08463 |
| CO2-Xe-131 | 9 | 1 | 8 | 9 | 0 | 9 | 2348.1034 | 2348.1035 | -0.0001 | 0.06856 |
| CO2-Xe-134 | 9 | 1 | 8 | 9 | 0 | 9 | 2348.1034 | 2348.1027 | 0.0007 | 0.03407 |
| CO2-Xe-136 | 9 | 1 | 8 | 9 | 0 | 9 | 2348.1034 | 2348.1022 | 0.0012 | 0.02906 |
| | | | | Blend | | | | 2348.1034 | 0.0000 |
| CO2-Xe-132 | 10 | 1 | 9 | 10 | 0 | 10 | 2348.1209 | 2348.1207 | 0.0002 | 0.05822 |
| CO2-Xe-129 | 10 | 1 | 9 | 10 | 0 | 10 | 2348.1209 | 2348.1217 | -0.0008 | 0.05622 |
| CO2-Xe-131 | 10 | 1 | 9 | 10 | 0 | 10 | 2348.1209 | 2348.1210 | -0.0001 | 0.04564 |
| CO2-Xe-134 | 10 | 1 | 9 | 10 | 0 | 10 | 2348.1209 | 2348.1200 | 0.0009 | 0.02274 |

| | | | | | | | | | |
|---|---|---|---|---|---|---|---|---|---|
| CO2-Xe-136 | 10 | 1 | 9 | 10 | 0 | 10 | 2348.1209 | 2348.1194 | 0.0015 | 0.01944 |
| | | | | | | | Blend | 2348.1209 | 0.0001 | |
| CO2-Xe-132 | 11 | 1 | 10 | 11 | 0 | 11 | 2348.1407 | 2348.1404 | 0.0003 | 0.03655 |
| CO2-Xe-129 | 11 | 1 | 10 | 11 | 0 | 11 | 2348.1407 | 2348.1417 | -0.0009 | 0.03518 |
| CO2-Xe-131 | 11 | 1 | 10 | 11 | 0 | 11 | 2348.1407 | 2348.1408 | -0.0001 | 0.02862 |
| CO2-Xe-134 | 11 | 1 | 10 | 11 | 0 | 11 | 2348.1407 | 2348.1396 | 0.0012 | 0.01431 |
| | | | | | | | Blend | 2348.1408 | -0.0001 | |
| CO2-Xe-132 | 2 | 1 | 2 | 1 | 0 | 1 | 2348.1610 | 2348.1607 | 0.0003 | 0.12812 |
| CO2-Xe-129 | 2 | 1 | 2 | 1 | 0 | 1 | 2348.1610 | 2348.1614 | -0.0004 | 0.12571 |
| CO2-Xe-131 | 2 | 1 | 2 | 1 | 0 | 1 | 2348.1610 | 2348.1609 | 0.0000 | 0.10096 |
| CO2-Xe-134 | 2 | 1 | 2 | 1 | 0 | 1 | 2348.1610 | 2348.1603 | 0.0007 | 0.04954 |
| CO2-Xe-136 | 2 | 1 | 2 | 1 | 0 | 1 | 2348.1610 | 2348.1598 | 0.0012 | 0.04193 |
| | | | | | | | Blend | 2348.1608 | 0.0002 | |
| CO2-Xe-132 | 3 | 1 | 3 | 2 | 0 | 2 | 2348.2238 | 2348.2237 | 0.0002 | 0.15534 |
| CO2-Xe-129 | 3 | 1 | 3 | 2 | 0 | 2 | 2348.2238 | 2348.2247 | -0.0008 | 0.15233 |
| CO2-Xe-131 | 3 | 1 | 3 | 2 | 0 | 2 | 2348.2238 | 2348.2240 | -0.0001 | 0.12239 |
| CO2-Xe-134 | 3 | 1 | 3 | 2 | 0 | 2 | 2348.2238 | 2348.2230 | 0.0009 | 0.06009 |
| CO2-Xe-136 | 3 | 1 | 3 | 2 | 0 | 2 | 2348.2238 | 2348.2223 | 0.0015 | 0.05087 |
| | | | | | | | Blend | 2348.2238 | 0.0000 | |
| CO2-Xe-132 | 4 | 1 | 4 | 3 | 0 | 3 | 2348.2853 | 2348.2851 | 0.0002 | 0.16855 |
| CO2-Xe-129 | 4 | 1 | 4 | 3 | 0 | 3 | 2348.2853 | 2348.2864 | -0.0012 | 0.16516 |
| CO2-Xe-131 | 4 | 1 | 4 | 3 | 0 | 3 | 2348.2853 | 2348.2855 | -0.0003 | 0.13276 |
| CO2-Xe-134 | 4 | 1 | 4 | 3 | 0 | 3 | 2348.2853 | 2348.2842 | 0.0010 | 0.06523 |
| CO2-Xe-136 | 4 | 1 | 4 | 3 | 0 | 3 | 2348.2853 | 2348.2834 | 0.0019 | 0.05525 |
| | | | | | | | Blend | 2348.2853 | 0.0000 | |
| CO2-Xe-132 | 5 | 1 | 5 | 4 | 0 | 4 | 2348.3454 | 2348.3450 | 0.0004 | 0.16759 |
| CO2-Xe-129 | 5 | 1 | 5 | 4 | 0 | 4 | 2348.3454 | 2348.3467 | -0.0012 | 0.16406 |
| CO2-Xe-131 | 5 | 1 | 5 | 4 | 0 | 4 | 2348.3454 | 2348.3456 | -0.0001 | 0.13197 |
| CO2-Xe-134 | 5 | 1 | 5 | 4 | 0 | 4 | 2348.3454 | 2348.3440 | 0.0015 | 0.0649 |
| | | | | | | | Blend | 2348.3455 | -0.0001 | |
| CO2-Xe-132 | 6 | 1 | 6 | 5 | 0 | 5 | 2348.4040 | 2348.4036 | 0.0005 | 0.15466 |
| CO2-Xe-129 | 6 | 1 | 6 | 5 | 0 | 5 | 2348.4040 | 2348.4055 | -0.0015 | 0.15121 |
| CO2-Xe-131 | 6 | 1 | 6 | 5 | 0 | 5 | 2348.4040 | 2348.4042 | -0.0002 | 0.12174 |
| CO2-Xe-134 | 6 | 1 | 6 | 5 | 0 | 5 | 2348.4040 | 2348.4023 | 0.0017 | 0.05994 |
| | | | | | | | Blend | 2348.4042 | -0.0001 | |
| CO2-Xe-132 | 7 | 1 | 7 | 6 | 0 | 6 | 2348.4614 | 2348.4607 | 0.0006 | 0.13347 |
| CO2-Xe-129 | 7 | 1 | 7 | 6 | 0 | 6 | 2348.4614 | 2348.4629 | -0.0016 | 0.1303 |
| CO2-Xe-131 | 7 | 1 | 7 | 6 | 0 | 6 | 2348.4614 | 2348.4614 | -0.0001 | 0.10501 |
| | | | | | | | Blend | 2348.4617 | -0.0004 | |
| CO2-Xe-132 | 8 | 1 | 8 | 7 | 0 | 7 | 2348.5173 | 2348.5166 | 0.0007 | 0.10825 |
| CO2-Xe-129 | 8 | 1 | 8 | 7 | 0 | 7 | 2348.5173 | 2348.5191 | -0.0017 | 0.10549 |
| CO2-Xe-131 | 8 | 1 | 8 | 7 | 0 | 7 | 2348.5173 | 2348.5174 | -0.0001 | 0.08511 |
| CO2-Xe-134 | 8 | 1 | 8 | 7 | 0 | 7 | 2348.5173 | 2348.5150 | 0.0023 | 0.04204 |
| | | | | | | | Blend | 2348.5174 | 0.0000 | |
| CO2-Xe-132 | 9 | 1 | 9 | 8 | 0 | 8 | 2348.5719 | 2348.5713 | 0.0006 | 0.08277 |
| CO2-Xe-131 | 9 | 1 | 9 | 8 | 0 | 8 | 2348.5719 | 2348.5722 | -0.0003 | 0.06504 |
| | | | | | | | Blend | 2348.5717 | 0.0002 | |
| CO2-Xe-132 | 10 | 1 | 10 | 9 | 0 | 9 | 2348.6255 | 2348.6249 | 0.0007 | 0.05981 |

| | | | | | | | | | |
|---|---|---|---|---|---|---|---|---|---|
| CO2-Xe-131 | 10 | 1 | 10 | 9 | 0 | 9 | 2348.6255 | 2348.6259 | -0.0003 | 0.04696 |
| | | | | | | Blend | | 2348.6253 | 0.0002 | |
| CO2-Xe-134 | 13 | 1 | 13 | 12 | 0 | 12 | 2348.7761 | 2348.7780 | -0.0019 | 0.00639 |
| CO2-Xe-136 | 13 | 1 | 13 | 12 | 0 | 12 | 2348.7761 | 2348.7757 | 0.0004 | 0.00548 |
| CO2-Xe-129 | 9 | 3 | 6 | 10 | 2 | 9 | 2348.7761 | 2348.7757 | 0.0003 | 0.00294 |
| | | | | | | Blend | | 2348.7767 | -0.0006 | |
| CO2-Xe-132 | 13 | 1 | 13 | 12 | 0 | 12 | 2348.7812 | 2348.7804 | 0.0009 | 0.01633 |
| CO2-Xe-131 | 13 | 1 | 13 | 12 | 0 | 12 | 2348.7812 | 2348.7816 | -0.0003 | 0.01279 |
| | | | | | | Blend | | 2348.7809 | 0.0003 | |
| CO2-Xe-129 | 13 | 1 | 13 | 12 | 0 | 12 | 2348.7836 | 2348.7840 | -0.0004 | |
| CO2-Xe-132 | 5 | 3 | 3 | 5 | 2 | 4 | 2349.4531 | 2349.4535 | -0.0004 | 0.03387 |
| CO2-Xe-132 | 5 | 3 | 2 | 5 | 2 | 3 | 2349.4531 | 2349.4528 | 0.0003 | 0.03386 |
| CO2-Xe-129 | 5 | 3 | 3 | 5 | 2 | 4 | 2349.4531 | 2349.4528 | 0.0003 | 0.03312 |
| CO2-Xe-129 | 5 | 3 | 2 | 5 | 2 | 3 | 2349.4531 | 2349.4521 | 0.0010 | 0.03312 |
| CO2-Xe-132 | 6 | 3 | 4 | 6 | 2 | 5 | 2349.4531 | 2349.4535 | -0.0004 | 0.03211 |
| CO2-Xe-132 | 6 | 3 | 3 | 6 | 2 | 4 | 2349.4531 | 2349.4522 | 0.0009 | 0.0321 |
| CO2-Xe-129 | 6 | 3 | 4 | 6 | 2 | 5 | 2349.4531 | 2349.4528 | 0.0003 | 0.03136 |
| CO2-Xe-129 | 6 | 3 | 3 | 6 | 2 | 4 | 2349.4531 | 2349.4514 | 0.0017 | 0.03134 |
| CO2-Xe-132 | 4 | 3 | 2 | 4 | 2 | 3 | 2349.4531 | 2349.4536 | -0.0004 | 0.03085 |
| CO2-Xe-132 | 4 | 3 | 1 | 4 | 2 | 2 | 2349.4531 | 2349.4533 | -0.0002 | 0.03085 |
| | | | | | | Blend | | 2349.4528 | 0.0003 | |
| CO2-Xe-132 | 3 | 3 | 0 | 2 | 2 | 1 | 2349.6561 | 2349.6561 | 0.0000 | 0.06827 |
| CO2-Xe-132 | 3 | 3 | 1 | 2 | 2 | 0 | 2349.6561 | 2349.6561 | 0.0000 | 0.06827 |
| CO2-Xe-129 | 3 | 3 | 0 | 2 | 2 | 1 | 2349.6561 | 2349.6565 | -0.0004 | 0.06698 |
| CO2-Xe-129 | 3 | 3 | 1 | 2 | 2 | 0 | 2349.6561 | 2349.6565 | -0.0003 | 0.06698 |
| CO2-Xe-131 | 3 | 3 | 0 | 2 | 2 | 1 | 2349.6561 | 2349.6563 | -0.0001 | 0.0538 |
| CO2-Xe-131 | 3 | 3 | 1 | 2 | 2 | 0 | 2349.6561 | 2349.6562 | -0.0001 | 0.0538 |
| CO2-Xe-134 | 3 | 3 | 0 | 2 | 2 | 1 | 2349.6561 | 2349.6559 | 0.0003 | 0.0264 |
| CO2-Xe-134 | 3 | 3 | 1 | 2 | 2 | 0 | 2349.6561 | 2349.6559 | 0.0003 | 0.0264 |
| CO2-Xe-136 | 3 | 3 | 0 | 2 | 2 | 1 | 2349.6561 | 2349.6556 | 0.0005 | 0.02234 |
| CO2-Xe-136 | 3 | 3 | 1 | 2 | 2 | 0 | 2349.6561 | 2349.6556 | 0.0005 | 0.02234 |
| | | | | | | Blend | | 2349.6562 | 0.0000 | |
| CO2-Xe-132 | 4 | 3 | 1 | 3 | 2 | 2 | 2349.7236 | 2349.7235 | 0.0001 | 0.06183 |
| CO2-Xe-132 | 4 | 3 | 2 | 3 | 2 | 1 | 2349.7236 | 2349.7234 | 0.0001 | 0.06183 |
| CO2-Xe-129 | 4 | 3 | 1 | 3 | 2 | 2 | 2349.7236 | 2349.7243 | -0.0007 | 0.06061 |
| CO2-Xe-129 | 4 | 3 | 2 | 3 | 2 | 1 | 2349.7236 | 2349.7242 | -0.0006 | 0.06061 |
| CO2-Xe-131 | 4 | 3 | 1 | 3 | 2 | 2 | 2349.7236 | 2349.7238 | -0.0002 | 0.04871 |
| CO2-Xe-131 | 4 | 3 | 2 | 3 | 2 | 1 | 2349.7236 | 2349.7237 | -0.0001 | 0.04871 |
| CO2-Xe-134 | 4 | 3 | 1 | 3 | 2 | 2 | 2349.7236 | 2349.7230 | 0.0005 | 0.02392 |
| CO2-Xe-134 | 4 | 3 | 2 | 3 | 2 | 1 | 2349.7236 | 2349.7230 | 0.0006 | 0.02392 |
| CO2-Xe-136 | 4 | 3 | 1 | 3 | 2 | 2 | 2349.7236 | 2349.7226 | 0.0010 | 0.02026 |
| CO2-Xe-136 | 4 | 3 | 2 | 3 | 2 | 1 | 2349.7236 | 2349.7225 | 0.0011 | 0.02026 |
| | | | | | | Blend | | 2349.7236 | 0.0000 | |
| CO2-Xe-132 | 5 | 3 | 2 | 4 | 2 | 3 | 2349.7911 | 2349.7909 | 0.0002 | 0.05417 |
| CO2-Xe-132 | 5 | 3 | 3 | 4 | 2 | 2 | 2349.7911 | 2349.7907 | 0.0005 | 0.05417 |
| CO2-Xe-129 | 5 | 3 | 2 | 4 | 2 | 3 | 2349.7911 | 2349.7921 | -0.0009 | 0.05305 |

| | | | | | | | | | |
|---|---|---|---|---|---|---|---|---|---|
| CO2-Xe-129 | 5 | 3 | 3 | 4 | 2 | 2 | 2349.7911 | 2349.7918 | -0.0006 | 0.05304 |
| CO2-Xe-131 | 5 | 3 | 2 | 4 | 2 | 3 | 2349.7911 | 2349.7913 | -0.0002 | 0.04266 |
| CO2-Xe-131 | 5 | 3 | 3 | 4 | 2 | 2 | 2349.7911 | 2349.7910 | 0.0001 | 0.04266 |
| CO2-Xe-134 | 5 | 3 | 2 | 4 | 2 | 3 | 2349.7911 | 2349.7902 | 0.0009 | 0.02097 |
| CO2-Xe-134 | 5 | 3 | 3 | 4 | 2 | 2 | 2349.7911 | 2349.7899 | 0.0012 | 0.02097 |
| CO2-Xe-136 | 5 | 3 | 2 | 4 | 2 | 3 | 2349.7911 | 2349.7895 | 0.0016 | 0.01777 |
| CO2-Xe-136 | 5 | 3 | 3 | 4 | 2 | 2 | 2349.7911 | 2349.7892 | 0.0019 | 0.01777 |
| | | | | | | Blend | | 2349.7910 | 0.0002 | |
| CO2-Xe-132 | 6 | 3 | 3 | 5 | 2 | 4 | 2349.8584 | 2349.8584 | 0.0001 | 0.0454 |
| CO2-Xe-132 | 6 | 3 | 4 | 5 | 2 | 3 | 2349.8584 | 2349.8577 | 0.0007 | 0.04539 |
| CO2-Xe-129 | 6 | 3 | 3 | 5 | 2 | 4 | 2349.8584 | 2349.8599 | -0.0014 | 0.04439 |
| CO2-Xe-129 | 6 | 3 | 4 | 5 | 2 | 3 | 2349.8584 | 2349.8592 | -0.0008 | 0.04438 |
| CO2-Xe-131 | 6 | 3 | 3 | 5 | 2 | 4 | 2349.8584 | 2349.8589 | -0.0004 | 0.03573 |
| CO2-Xe-131 | 6 | 3 | 4 | 5 | 2 | 3 | 2349.8584 | 2349.8582 | 0.0002 | 0.03573 |
| CO2-Xe-134 | 6 | 3 | 3 | 5 | 2 | 4 | 2349.8584 | 2349.8574 | 0.0010 | 0.01759 |
| CO2-Xe-134 | 6 | 3 | 4 | 5 | 2 | 3 | 2349.8584 | 2349.8567 | 0.0017 | 0.01759 |
| CO2-Xe-136 | 6 | 3 | 3 | 5 | 2 | 4 | 2349.8584 | 2349.8565 | 0.0020 | 0.01492 |
| | | | | | | Blend | | 2349.8584 | 0.0000 | |

Table A-4. Observed and fitted combination band transitions of CO$_2$-Xe (cm$^{-1}$).

Int = calculated relative intensity, used to weight the blended lines.

Transitions with $Ka'$ = 0 and 2 belong to the bend combination band.

Transitions with $Ka'$ = 1 belong to the stretch combination band.

| Isotope | $J'$ | $Ka'$ | $Kc'$ | $J''$ | $Ka''$ | $Kc''$ | Observed | Calculated | Obs - Calc | Int |
|---------|------|-------|-------|-------|--------|--------|----------|------------|------------|-----|
| CO2-Xe-132 | 10 | 2 | 9 | 11 | 2 | 10 | 2377.3332 | 2377.3325 | 0.0007 | |
| CO2-Xe-131 | 10 | 2 | 9 | 11 | 2 | 10 | 2377.3395 | 2377.3393 | 0.0002 | |
| CO2-Xe-129 | 10 | 2 | 9 | 11 | 2 | 10 | 2377.3503 | 2377.3507 | -0.0003 | |
| CO2-Xe-132 | 10 | 0 | 10 | 11 | 0 | 11 | 2377.3806 | 2377.3810 | -0.0004 | 0.05252 |
| CO2-Xe-129 | 10 | 0 | 10 | 11 | 0 | 11 | 2377.3806 | 2377.3798 | 0.0008 | 0.05083 |
| CO2-Xe-131 | 10 | 0 | 10 | 11 | 0 | 11 | 2377.3806 | 2377.3806 | 0.0000 | 0.04120 |
| CO2-Xe-134 | 10 | 0 | 10 | 11 | 0 | 11 | 2377.3806 | 2377.3816 | -0.0010 | 0.02048 |
| CO2-Xe-136 | 10 | 0 | 10 | 11 | 0 | 11 | 2377.3806 | 2377.3821 | -0.0015 | 0.01749 |
| | | | | | | | Blend | 2377.3808 | -0.0002 | |
| CO2-Xe-131 | 9 | 1 | 9 | 10 | 2 | 8 | 2377.3910 | 2377.3917 | -0.0007 | |
| CO2-Xe-129 | 9 | 2 | 7 | 10 | 2 | 8 | 2377.4121 | 2377.4117 | 0.0005 | |
| CO2-Xe-132 | 9 | 2 | 8 | 10 | 2 | 9 | 2377.4309 | 2377.4308 | 0.0001 | |
| CO2-Xe-131 | 9 | 2 | 8 | 10 | 2 | 9 | 2377.4376 | 2377.4377 | -0.0001 | |
| CO2-Xe-129 | 9 | 2 | 8 | 10 | 2 | 9 | 2377.4486 | 2377.4491 | -0.0004 | |
| CO2-Xe-132 | 9 | 0 | 9 | 10 | 0 | 10 | 2377.4704 | 2377.4706 | -0.0002 | 0.07784 |
| CO2-Xe-129 | 9 | 0 | 9 | 10 | 0 | 10 | 2377.4704 | 2377.4696 | 0.0008 | 0.07550 |
| CO2-Xe-131 | 9 | 0 | 9 | 10 | 0 | 10 | 2377.4704 | 2377.4703 | 0.0001 | 0.06111 |
| CO2-Xe-134 | 9 | 0 | 9 | 10 | 0 | 10 | 2377.4704 | 2377.4712 | -0.0008 | 0.03032 |
| CO2-Xe-136 | 9 | 0 | 9 | 10 | 0 | 10 | 2377.4704 | 2377.4716 | -0.0012 | 0.02585 |
| | | | | | | | Blend | 2377.4704 | 0.0000 | |
| CO2-Xe-132 | 8 | 1 | 8 | 9 | 2 | 7 | 2377.4876 | 2377.4878 | -0.0002 | |
| CO2-Xe-131 | 8 | 2 | 6 | 9 | 2 | 7 | 2377.4988 | 2377.4990 | -0.0002 | |
| CO2-Xe-129 | 8 | 2 | 6 | 9 | 2 | 7 | 2377.5181 | 2377.5177 | 0.0004 | |
| CO2-Xe-132 | 8 | 2 | 7 | 9 | 2 | 8 | 2377.5272 | 2377.5273 | -0.0001 | |

| | | | | | | | | | |
|---|---|---|---|---|---|---|---|---|---|
| CO2-Xe-131 | 8 | 2 | 7 | 9 | 2 | 8 | 2377.5342 | 2377.5343 | -0.0001 | |
| | | | | | | | | | |
| CO2-Xe-129 | 8 | 2 | 7 | 9 | 2 | 8 | 2377.5450 | 2377.5455 | -0.0005 | |
| | | | | | | | | | |
| CO2-Xe-132 | 8 | 0 | 8 | 9 | 0 | 9 | 2377.5578 | 2377.5580 | -0.0002 | 0.10926 |
| CO2-Xe-129 | 8 | 0 | 8 | 9 | 0 | 9 | 2377.5578 | 2377.5571 | 0.0007 | 0.10619 |
| CO2-Xe-131 | 8 | 0 | 8 | 9 | 0 | 9 | 2377.5578 | 2377.5577 | 0.0001 | 0.08584 |
| CO2-Xe-134 | 8 | 0 | 8 | 9 | 0 | 9 | 2377.5578 | 2377.5584 | -0.0006 | 0.04250 |
| CO2-Xe-136 | 8 | 0 | 8 | 9 | 0 | 9 | 2377.5578 | 2377.5587 | -0.0010 | 0.03619 |
| | | | | | | | Blend | 2377.5578 | 0.0000 | |
| | | | | | | | | | |
| CO2-Xe-132 | 7 | 1 | 7 | 8 | 2 | 6 | 2377.5923 | 2377.5925 | -0.0002 | |
| | | | | | | | | | |
| CO2-Xe-131 | 7 | 2 | 5 | 8 | 2 | 6 | 2377.6036 | 2377.6031 | 0.0005 | 0.02308 |
| CO2-Xe-134 | 7 | 2 | 6 | 8 | 2 | 7 | 2377.6036 | 2377.6049 | -0.0014 | 0.01162 |
| | | | | | | | Blend | 2377.6037 | -0.0001 | |
| CO2-Xe-132 | 7 | 2 | 6 | 8 | 2 | 7 | 2377.6215 | 2377.6222 | -0.0007 | 0.03517 |
| CO2-Xe-129 | 7 | 2 | 5 | 8 | 2 | 6 | 2377.6215 | 2377.6203 | 0.0012 | 0.03376 |
| | | | | | | | Blend | 2377.6213 | 0.0002 | |
| CO2-Xe-131 | 7 | 2 | 6 | 8 | 2 | 7 | 2377.6289 | 2377.6291 | -0.0002 | 0.02948 |
| CO2-Xe-134 | 9 | 2 | 7 | 10 | 2 | 8 | 2377.6289 | 2377.6279 | 0.0010 | 0.00737 |
| | | | | | | | Blend | 2377.6289 | 0.0000 | |
| CO2-Xe-132 | 7 | 0 | 7 | 8 | 0 | 8 | 2377.6428 | 2377.6430 | -0.0002 | 0.14489 |
| CO2-Xe-129 | 7 | 0 | 7 | 8 | 0 | 8 | 2377.6428 | 2377.6424 | 0.0004 | 0.14109 |
| CO2-Xe-131 | 7 | 0 | 7 | 8 | 0 | 8 | 2377.6428 | 2377.6428 | 0.0000 | 0.11390 |
| CO2-Xe-134 | 7 | 0 | 7 | 8 | 0 | 8 | 2377.6428 | 2377.6433 | -0.0005 | 0.05630 |
| CO2-Xe-136 | 7 | 0 | 7 | 8 | 0 | 8 | 2377.6428 | 2377.6436 | -0.0008 | 0.04787 |
| | | | | | | | Blend | 2377.6429 | -0.0001 | |
| CO2-Xe-132 | 6 | 2 | 4 | 7 | 2 | 5 | 2377.6941 | 2377.6941 | 0.0000 | 0.03466 |
| CO2-Xe-136 | 8 | 2 | 6 | 9 | 2 | 7 | 2377.6941 | 2377.6920 | 0.0021 | 0.00954 |
| | | | | | | | Blend | 2377.6937 | 0.0005 | |
| | | | | | | | | | |
| CO2-Xe-134 | 6 | 2 | 5 | 7 | 2 | 6 | 2377.6979 | 2377.6979 | 0.0000 | |
| | | | | | | | | | |
| CO2-Xe-131 | 6 | 2 | 4 | 7 | 2 | 5 | 2377.7038 | 2377.7040 | -0.0002 | |
| | | | | | | | | | |
| CO2-Xe-132 | 6 | 2 | 5 | 7 | 2 | 6 | 2377.7152 | 2377.7154 | -0.0002 | |
| | | | | | | | | | |
| CO2-Xe-132 | 6 | 0 | 6 | 7 | 0 | 7 | 2377.7257 | 2377.7259 | -0.0002 | 0.18092 |
| CO2-Xe-129 | 6 | 0 | 6 | 7 | 0 | 7 | 2377.7257 | 2377.7255 | 0.0002 | 0.17647 |
| CO2-Xe-131 | 6 | 0 | 6 | 7 | 0 | 7 | 2377.7257 | 2377.7258 | -0.0001 | 0.14230 |
| CO2-Xe-134 | 6 | 0 | 6 | 7 | 0 | 7 | 2377.7257 | 2377.7260 | -0.0004 | 0.07022 |
| CO2-Xe-136 | 6 | 0 | 6 | 7 | 0 | 7 | 2377.7257 | 2377.7261 | -0.0005 | 0.05965 |
| | | | | | | | Blend | 2377.7258 | -0.0001 | |
| | | | | | | | | | |
| CO2-Xe-129 | 6 | 2 | 5 | 7 | 2 | 6 | 2377.7324 | 2377.7326 | -0.0002 | |
| | | | | | | | | | |
| CO2-Xe-131 | 8 | 1 | 8 | 9 | 2 | 7 | 2377.7385 | 2377.7384 | 0.0001 | |

| | | | | | | | | | |
|---|---|---|---|---|---|---|---|---|---|
| CO2-Xe-132 | 5 | 2 | 3 | 6 | 2 | 4 | 2377.7929 | 2377.7927 | 0.0002 | |
| | | | | | | | | | | |
| CO2-Xe-132 | 5 | 0 | 5 | 6 | 0 | 6 | 2377.8063 | 2377.8065 | -0.0002 | 0.21168 |
| CO2-Xe-129 | 5 | 0 | 5 | 6 | 0 | 6 | 2377.8063 | 2377.8064 | -0.0001 | 0.20678 |
| CO2-Xe-131 | 5 | 0 | 5 | 6 | 0 | 6 | 2377.8063 | 2377.8065 | -0.0002 | 0.16657 |
| CO2-Xe-134 | 5 | 0 | 5 | 6 | 0 | 6 | 2377.8063 | 2377.8065 | -0.0002 | 0.08208 |
| CO2-Xe-136 | 5 | 0 | 5 | 6 | 0 | 6 | 2377.8063 | 2377.8065 | -0.0001 | 0.06966 |
| | | | | | | | Blend | 2377.8065 | -0.0001 | |
| CO2-Xe-129 | 5 | 2 | 3 | 6 | 2 | 4 | 2377.8145 | 2377.8150 | -0.0005 | 0.05449 |
| CO2-Xe-131 | 5 | 2 | 4 | 6 | 2 | 5 | 2377.8145 | 2377.8134 | 0.0011 | 0.04313 |
| | | | | | | | Blend | 2377.8143 | 0.0002 | |
| | | | | | | | | | | |
| CO2-Xe-129 | 5 | 2 | 4 | 6 | 2 | 5 | 2377.8229 | 2377.8229 | 0.0000 | |
| | | | | | | | | | | |
| CO2-Xe-134 | 4 | 1 | 4 | 5 | 2 | 3 | 2377.8662 | 2377.8662 | 0.0000 | 0.01307 |
| CO2-Xe-136 | 6 | 2 | 5 | 7 | 2 | 6 | 2377.8662 | 2377.8654 | 0.0008 | 0.01204 |
| | | | | | | | Blend | 2377.8658 | 0.0004 | |
| CO2-Xe-132 | 6 | 1 | 6 | 7 | 2 | 5 | 2377.8786 | 2377.8787 | -0.0001 | 0.03010 |
| CO2-Xe-134 | 4 | 2 | 3 | 5 | 2 | 4 | 2377.8786 | 2377.8796 | -0.0010 | 0.01622 |
| | | | | | | | Blend | 2377.8790 | -0.0004 | |
| CO2-Xe-132 | 4 | 0 | 4 | 5 | 0 | 5 | 2377.8849 | 2377.8850 | -0.0001 | 0.23032 |
| CO2-Xe-129 | 4 | 0 | 4 | 5 | 0 | 5 | 2377.8849 | 2377.8852 | -0.0003 | 0.22527 |
| CO2-Xe-131 | 4 | 0 | 4 | 5 | 0 | 5 | 2377.8849 | 2377.8850 | -0.0002 | 0.18131 |
| CO2-Xe-134 | 4 | 0 | 4 | 5 | 0 | 5 | 2377.8849 | 2377.8848 | 0.0001 | 0.08923 |
| CO2-Xe-136 | 4 | 0 | 4 | 5 | 0 | 5 | 2377.8849 | 2377.8846 | 0.0003 | 0.07567 |
| | | | | | | | Blend | 2377.8850 | -0.0001 | |
| CO2-Xe-132 | 4 | 2 | 3 | 5 | 2 | 4 | 2377.8964 | 2377.8968 | -0.0005 | 0.05421 |
| CO2-Xe-131 | 4 | 2 | 2 | 5 | 2 | 3 | 2377.8964 | 2377.8961 | 0.0003 | 0.04227 |
| | | | | | | | Blend | 2377.8965 | -0.0001 | |
| | | | | | | | | | | |
| CO2-Xe-131 | 4 | 2 | 3 | 5 | 2 | 4 | 2377.9029 | 2377.9028 | 0.0001 | |
| | | | | | | | | | | |
| CO2-Xe-129 | 4 | 2 | 2 | 5 | 2 | 3 | 2377.9071 | 2377.9068 | 0.0003 | |
| | | | | | | | | | | |
| CO2-Xe-129 | 4 | 2 | 3 | 5 | 2 | 4 | 2377.9111 | 2377.9110 | 0.0000 | |
| | | | | | | | | | | |
| CO2-Xe-132 | 5 | 1 | 5 | 6 | 2 | 4 | 2377.9497 | 2377.9499 | -0.0002 | 0.03082 |
| CO2-Xe-134 | 5 | 1 | 4 | 6 | 2 | 5 | 2377.9497 | 2377.9492 | 0.0005 | 0.01259 |
| | | | | | | | Blend | 2377.9497 | 0.0000 | |
| CO2-Xe-132 | 3 | 0 | 3 | 4 | 0 | 4 | 2377.9612 | 2377.9613 | -0.0001 | 0.23011 |
| CO2-Xe-129 | 3 | 0 | 3 | 4 | 0 | 4 | 2377.9612 | 2377.9618 | -0.0006 | 0.22532 |
| CO2-Xe-131 | 3 | 0 | 3 | 4 | 0 | 4 | 2377.9612 | 2377.9614 | -0.0003 | 0.18121 |
| CO2-Xe-134 | 3 | 0 | 3 | 4 | 0 | 4 | 2377.9612 | 2377.9609 | 0.0003 | 0.08909 |
| CO2-Xe-136 | 3 | 0 | 3 | 4 | 0 | 4 | 2377.9612 | 2377.9605 | 0.0007 | 0.07549 |
| | | | | | | | Blend | 2377.9613 | -0.0001 | |
| | | | | | | | | | | |
| CO2-Xe-132 | 3 | 2 | 1 | 4 | 2 | 2 | 2377.9805 | 2377.9808 | -0.0003 | |

| | | | | | | | | | |
|---|---|---|---|---|---|---|---|---|---|
| CO2-Xe-131 | 3 | 2 | 2 | 4 | 2 | 3 | 2377.9898 | 2377.9901 | -0.0003 | |
| | | | | | | | | | | |
| CO2-Xe-129 | 3 | 2 | 1 | 4 | 2 | 2 | 2377.9956 | 2377.9947 | 0.0009 | 0.05814 |
| CO2-Xe-129 | 3 | 2 | 2 | 4 | 2 | 3 | 2377.9956 | 2377.9964 | -0.0008 | 0.05939 |
| | | | | | | | Blend | 2377.9956 | 0.0000 | |
| | | | | | | | | | | |
| CO2-Xe-132 | 4 | 1 | 4 | 5 | 2 | 3 | 2378.0184 | 2378.0181 | 0.0003 | |
| | | | | | | | | | | |
| CO2-Xe-132 | 2 | 0 | 2 | 3 | 0 | 3 | 2378.0356 | 2378.0354 | 0.0002 | 0.20615 |
| CO2-Xe-129 | 2 | 0 | 2 | 3 | 0 | 3 | 2378.0356 | 2378.0362 | -0.0006 | 0.20204 |
| CO2-Xe-134 | 2 | 0 | 2 | 3 | 0 | 3 | 2378.0356 | 2378.0348 | 0.0008 | 0.07977 |
| CO2-Xe-136 | 2 | 0 | 2 | 3 | 0 | 3 | 2378.0356 | 2378.0342 | 0.0014 | 0.06756 |
| CO2-Xe-131 | 2 | 0 | 2 | 3 | 0 | 3 | 2378.0356 | 2378.0357 | -0.0001 | 0.16239 |
| | | | | | | | Blend | 2378.0355 | 0.0001 | |
| CO2-Xe-132 | 2 | 2 | 1 | 3 | 2 | 2 | 2378.0707 | 2378.0716 | -0.0009 | 0.03757 |
| CO2-Xe-132 | 2 | 2 | 0 | 3 | 2 | 1 | 2378.0707 | 2378.0701 | 0.0005 | 0.03631 |
| | | | | | | | Blend | 2378.0709 | -0.0002 | |
| CO2-Xe-131 | 2 | 2 | 0 | 3 | 2 | 1 | 2378.0742 | 2378.0740 | 0.0002 | 0.03110 |
| CO2-Xe-131 | 2 | 2 | 1 | 3 | 2 | 2 | 2378.0742 | 2378.0749 | -0.0007 | 0.03163 |
| | | | | | | | Blend | 2378.0744 | -0.0002 | |
| CO2-Xe-129 | 2 | 2 | 1 | 3 | 2 | 2 | 2378.0787 | 2378.0787 | 0.0000 | 0.04154 |
| CO2-Xe-129 | 2 | 2 | 0 | 3 | 2 | 1 | 2378.0787 | 2378.0783 | 0.0004 | 0.04130 |
| | | | | | | | Blend | 2378.0785 | 0.0002 | |
| CO2-Xe-129 | 1 | 0 | 1 | 2 | 0 | 2 | 2378.1077 | 2378.1085 | -0.0008 | 0.15399 |
| CO2-Xe-131 | 1 | 0 | 1 | 2 | 0 | 2 | 2378.1077 | 2378.1077 | 0.0000 | 0.12372 |
| CO2-Xe-134 | 1 | 0 | 1 | 2 | 0 | 2 | 2378.1077 | 2378.1065 | 0.0012 | 0.06073 |
| CO2-Xe-132 | 1 | 0 | 1 | 2 | 0 | 2 | 2378.1077 | 2378.1073 | 0.0004 | 0.15702 |
| | | | | | | | Blend | 2378.1077 | 0.0000 | |
| CO2-Xe-132 | 6 | 2 | 4 | 6 | 2 | 5 | 2378.1674 | 2378.1687 | -0.0014 | 0.00904 |
| CO2-Xe-129 | 7 | 2 | 5 | 7 | 2 | 6 | 2378.1674 | 2378.1669 | 0.0005 | 0.00685 |
| CO2-Xe-131 | 7 | 2 | 6 | 7 | 2 | 5 | 2378.1674 | 2378.1672 | 0.0001 | 0.00595 |
| CO2-Xe-131 | 2 | 1 | 2 | 3 | 2 | 1 | 2378.1674 | 2378.1671 | 0.0003 | 0.00402 |
| | | | | | | | Blend | 2378.1676 | -0.0003 | |
| CO2-Xe-132 | 0 | 0 | 0 | 1 | 0 | 1 | 2378.1776 | 2378.1770 | 0.0006 | 0.08579 |
| CO2-Xe-129 | 0 | 0 | 0 | 1 | 0 | 1 | 2378.1776 | 2378.1786 | -0.0010 | 0.08418 |
| CO2-Xe-131 | 0 | 0 | 0 | 1 | 0 | 1 | 2378.1776 | 2378.1776 | 0.0000 | 0.06761 |
| CO2-Xe-134 | 0 | 0 | 0 | 1 | 0 | 1 | 2378.1776 | 2378.1760 | 0.0016 | 0.03317 |
| | | | | | | | Blend | 2378.1775 | 0.0001 | |
| | | | | | | | | | | |
| CO2-Xe-132 | 4 | 2 | 2 | 4 | 2 | 3 | 2378.2260 | 2378.2263 | -0.0003 | |
| | | | | | | | | | | |
| CO2-Xe-129 | 5 | 2 | 4 | 5 | 2 | 3 | 2378.2294 | 2378.2294 | 0.0000 | |
| | | | | | | | | | | |
| CO2-Xe-132 | 4 | 2 | 3 | 4 | 2 | 2 | 2378.2343 | 2378.2340 | 0.0003 | 0.02896 |
| CO2-Xe-131 | 4 | 2 | 2 | 4 | 2 | 3 | 2378.2343 | 2378.2348 | -0.0004 | 0.02256 |
| | | | | | | | Blend | 2378.2343 | 0.0000 | |
| CO2-Xe-131 | 4 | 2 | 3 | 4 | 2 | 2 | 2378.2405 | 2378.2405 | 0.0000 | |
| CO2-Xe-129 | 4 | 2 | 2 | 4 | 2 | 3 | 2378.2471 | 2378.2468 | 0.0003 | |
| CO2-Xe-132 | 3 | 2 | 1 | 3 | 2 | 2 | 2378.2506 | 2378.2510 | -0.0004 | 0.04399 |
| CO2-Xe-129 | 4 | 2 | 3 | 4 | 2 | 2 | 2378.2506 | 2378.2500 | 0.0006 | 0.03392 |
| | | | | | | | Blend | 2378.2506 | 0.0000 | |
| CO2-Xe-132 | 3 | 2 | 2 | 3 | 2 | 1 | 2378.2549 | 2378.2550 | -0.0001 | |
| CO2-Xe-129 | 3 | 2 | 1 | 3 | 2 | 2 | 2378.2674 | 2378.2665 | 0.0009 | 0.05398 |
| CO2-Xe-129 | 3 | 2 | 2 | 3 | 2 | 1 | 2378.2674 | 2378.2678 | -0.0005 | 0.05516 |
| | | | | | | | Blend | 2378.2672 | 0.0002 | |
| CO2-Xe-132 | 2 | 2 | 1 | 2 | 2 | 0 | 2378.2732 | 2378.2741 | -0.0008 | 0.08578 |
| CO2-Xe-132 | 2 | 2 | 0 | 2 | 2 | 1 | 2378.2732 | 2378.2727 | 0.0005 | 0.08289 |
| | | | | | | | Blend | 2378.2734 | -0.0002 | |
| CO2-Xe-131 | 2 | 2 | 1 | 2 | 2 | 0 | 2378.2772 | 2378.2777 | -0.0005 | 0.07223 |
| CO2-Xe-131 | 2 | 2 | 0 | 2 | 2 | 1 | 2378.2772 | 2378.2769 | 0.0003 | 0.07100 |
| | | | | | | | Blend | 2378.2774 | -0.0001 | |
| CO2-Xe-129 | 2 | 2 | 0 | 2 | 2 | 1 | 2378.2823 | 2378.2820 | 0.0003 | 0.09437 |
| CO2-Xe-129 | 2 | 2 | 1 | 2 | 2 | 0 | 2378.2823 | 2378.2823 | 0.0000 | 0.09491 |
| | | | | | | | Blend | 2378.2822 | 0.0001 | |
| CO2-Xe-136 | 2 | 2 | 1 | 2 | 2 | 0 | 2378.2971 | 2378.2999 | -0.0028 | 0.02558 |
| CO2-Xe-136 | 2 | 2 | 0 | 2 | 2 | 1 | 2378.2971 | 2378.2982 | -0.0011 | 0.02685 |
| | | | | | | | Blend | 2378.2990 | -0.0020 | |
| CO2-Xe-131 | 1 | 0 | 1 | 0 | 0 | 0 | 2378.3102 | 2378.3106 | -0.0004 | 0.07064 |
| CO2-Xe-132 | 1 | 0 | 1 | 0 | 0 | 0 | 2378.3102 | 2378.3099 | 0.0003 | 0.08963 |
| | | | | | | | Blend | 2378.3102 | 0.0000 | |
| CO2-Xe-132 | 2 | 1 | 2 | 2 | 2 | 1 | 2378.3497 | 2378.3487 | 0.0010 | |
| CO2-Xe-132 | 4 | 1 | 4 | 4 | 2 | 3 | 2378.3561 | 2378.3561 | 0.0000 | 0.01478 |
| CO2-Xe-132 | 5 | 1 | 5 | 5 | 2 | 4 | 2378.3561 | 2378.3560 | 0.0001 | 0.01112 |
| CO2-Xe-132 | 2 | 1 | 1 | 2 | 2 | 0 | 2378.3561 | 2378.3558 | 0.0003 | 0.01576 |
| | | | | | | | Blend | 2378.3560 | 0.0001 | |
| CO2-Xe-132 | 2 | 0 | 2 | 1 | 0 | 1 | 2378.3734 | 2378.3729 | 0.0005 | 0.17139 |
| CO2-Xe-131 | 2 | 0 | 2 | 1 | 0 | 1 | 2378.3734 | 2378.3738 | -0.0004 | 0.13507 |
| | | | | | | | Blend | 2378.3733 | 0.0001 | |
| CO2-Xe-129 | 2 | 0 | 2 | 1 | 0 | 1 | 2378.3756 | 2378.3756 | 0.0000 | |
| CO2-Xe-130 | 4 | 4 | 0 | 4 | 4 | 1 | 2378.4176 | 2378.4177 | -0.0001 | 0.00143 |
| CO2-Xe-130 | 4 | 4 | 1 | 4 | 4 | 0 | 2378.4176 | 2378.4177 | -0.0001 | 0.00143 |
| CO2-Xe-132 | 4 | 4 | 0 | 4 | 4 | 1 | 2378.4176 | 2378.4164 | 0.0012 | 0.00939 |
| CO2-Xe-132 | 4 | 4 | 1 | 4 | 4 | 0 | 2378.4176 | 2378.4164 | 0.0012 | 0.00939 |
| CO2-Xe-131 | 4 | 4 | 0 | 4 | 4 | 1 | 2378.4176 | 2378.4171 | 0.0005 | 0.00740 |
| CO2-Xe-131 | 4 | 4 | 1 | 4 | 4 | 0 | 2378.4176 | 2378.4171 | 0.0005 | 0.00740 |
| CO2-Xe-129 | 4 | 4 | 0 | 4 | 4 | 1 | 2378.4176 | 2378.4184 | -0.0008 | 0.00921 |

| | | | | | | | | | |
|---|---|---|---|---|---|---|---|---|---|
| CO2-Xe-129 | 4 | 4 | 1 | 4 | 4 | 0 | 2378.4176 | 2378.4184 | -0.0008 | 0.00921 |
| | | | | | | | Blend | 2378.4173 | 0.0003 | |
| CO2-Xe-132 | 3 | 0 | 3 | 2 | 0 | 2 | 2378.4341 | 2378.4337 | 0.0004 | 0.23508 |
| CO2-Xe-131 | 3 | 0 | 3 | 2 | 0 | 2 | 2378.4341 | 2378.4347 | -0.0006 | 0.18523 |
| | | | | | | | Blend | 2378.4341 | -0.0001 | |
| CO2-Xe-129 | 3 | 0 | 3 | 2 | 0 | 2 | 2378.4370 | 2378.4368 | 0.0002 | |
| CO2-Xe-132 | 3 | 2 | 1 | 2 | 2 | 0 | 2378.4536 | 2378.4535 | 0.0001 | |
| CO2-Xe-132 | 3 | 2 | 2 | 2 | 2 | 1 | 2378.4574 | 2378.4576 | -0.0002 | |
| CO2-Xe-131 | 3 | 2 | 1 | 2 | 2 | 0 | 2378.4607 | 2378.4606 | 0.0001 | |
| CO2-Xe-131 | 3 | 2 | 2 | 2 | 2 | 1 | 2378.4632 | 2378.4634 | -0.0003 | |
| CO2-Xe-129 | 3 | 2 | 2 | 2 | 2 | 1 | 2378.4712 | 2378.4716 | -0.0003 | 0.04502 |
| CO2-Xe-129 | 3 | 2 | 1 | 2 | 2 | 0 | 2378.4712 | 2378.4701 | 0.0011 | 0.04407 |
| CO2-Xe-134 | 4 | 1 | 4 | 3 | 2 | 1 | 2378.4712 | 2378.4720 | -0.0008 | 0.01385 |
| | | | | | | | Blend | 2378.4710 | 0.0002 | |
| CO2-Xe-132 | 4 | 0 | 4 | 3 | 0 | 3 | 2378.4925 | 2378.4921 | 0.0004 | 0.27409 |
| CO2-Xe-131 | 4 | 0 | 4 | 3 | 0 | 3 | 2378.4925 | 2378.4933 | -0.0008 | 0.21593 |
| | | | | | | | Blend | 2378.4926 | -0.0001 | |
| CO2-Xe-129 | 4 | 0 | 4 | 3 | 0 | 3 | 2378.4959 | 2378.4956 | 0.0003 | |
| CO2-Xe-131 | 4 | 2 | 2 | 3 | 2 | 1 | 2378.5045 | 2378.5051 | -0.0007 | 0.04493 |
| CO2-Xe-132 | 4 | 2 | 3 | 3 | 2 | 2 | 2378.5045 | 2378.5043 | 0.0002 | 0.05762 |
| | | | | | | | Blend | 2378.5046 | -0.0002 | |
| CO2-Xe-134 | 5 | 1 | 5 | 4 | 2 | 2 | 2378.5112 | 2378.5099 | 0.0013 | 0.01548 |
| CO2-Xe-131 | 4 | 2 | 3 | 3 | 2 | 2 | 2378.5112 | 2378.5113 | -0.0001 | 0.04927 |
| | | | | | | | Blend | 2378.5110 | 0.0003 | |
| CO2-Xe-129 | 4 | 2 | 2 | 3 | 2 | 1 | 2378.5184 | 2378.5182 | 0.0002 | |
| CO2-Xe-129 | 4 | 2 | 3 | 3 | 2 | 2 | 2378.5220 | 2378.5218 | 0.0002 | |
| CO2-Xe-132 | 5 | 2 | 3 | 4 | 2 | 2 | 2378.5360 | 2378.5360 | -0.0001 | |
| CO2-Xe-136 | 5 | 0 | 5 | 4 | 0 | 4 | 2378.5428 | 2378.5429 | -0.0001 | |
| CO2-Xe-132 | 5 | 0 | 5 | 4 | 0 | 4 | 2378.5486 | 2378.5482 | 0.0005 | 0.28656 |
| CO2-Xe-131 | 5 | 0 | 5 | 4 | 0 | 4 | 2378.5486 | 2378.5495 | -0.0009 | 0.22569 |
| | | | | | | | Blend | 2378.5487 | -0.0001 | |
| CO2-Xe-129 | 5 | 0 | 5 | 4 | 0 | 4 | 2378.5522 | 2378.5521 | 0.0001 | |
| CO2-Xe-132 | 3 | 1 | 3 | 2 | 2 | 0 | 2378.5568 | 2378.5561 | 0.0007 | 0.01488 |

| | | | | | | | | | |
|---|---|---|---|---|---|---|---|---|---|
| CO2-Xe-131 | 5 | 2 | 4 | 4 | 2 | 3 | 2378.5568 | 2378.5570 | -0.0002 | 0.05525 |
| | | | | | | | Blend | 2378.5568 | 0.0000 | |
| CO2-Xe-129 | 5 | 2 | 3 | 4 | 2 | 2 | 2378.5629 | 2378.5624 | 0.0005 | |
| CO2-Xe-129 | 5 | 2 | 4 | 4 | 2 | 3 | 2378.5695 | 2378.5693 | 0.0001 | |
| CO2-Xe-134 | 6 | 2 | 5 | 5 | 2 | 4 | 2378.5729 | 2378.5719 | 0.0010 | 0.02071 |
| CO2-Xe-132 | 6 | 2 | 4 | 5 | 2 | 3 | 2378.5729 | 2378.5729 | 0.0000 | 0.05094 |
| | | | | | | | Blend | 2378.5726 | 0.0003 | |
| CO2-Xe-131 | 6 | 2 | 4 | 5 | 2 | 3 | 2378.5845 | 2378.5844 | 0.0001 | |
| CO2-Xe-132 | 6 | 2 | 5 | 5 | 2 | 4 | 2378.5925 | 2378.5925 | 0.0000 | |
| CO2-Xe-136 | 6 | 0 | 6 | 5 | 0 | 5 | 2378.5957 | 2378.5959 | -0.0003 | |
| CO2-Xe-131 | 6 | 0 | 6 | 5 | 0 | 5 | 2378.6024 | 2378.6033 | -0.0008 | 0.21657 |
| CO2-Xe-132 | 6 | 0 | 6 | 5 | 0 | 5 | 2378.6024 | 2378.6018 | 0.0007 | 0.27506 |
| | | | | | | | Blend | 2378.6024 | 0.0000 | |
| CO2-Xe-129 | 6 | 0 | 6 | 5 | 0 | 5 | 2378.6064 | 2378.6062 | 0.0001 | |
| CO2-Xe-129 | 6 | 2 | 5 | 5 | 2 | 4 | 2378.6141 | 2378.6145 | -0.0004 | |
| CO2-Xe-131 | 7 | 2 | 5 | 6 | 2 | 4 | 2378.6195 | 2378.6194 | 0.0001 | |
| CO2-Xe-132 | 4 | 1 | 4 | 3 | 2 | 1 | 2378.6265 | 2378.6260 | 0.0004 | |
| CO2-Xe-132 | 7 | 2 | 6 | 6 | 2 | 5 | 2378.6340 | 2378.6340 | 0.0000 | |
| CO2-Xe-129 | 7 | 2 | 5 | 6 | 2 | 4 | 2378.6419 | 2378.6404 | 0.0015 | 0.05655 |
| CO2-Xe-131 | 7 | 2 | 6 | 6 | 2 | 5 | 2378.6419 | 2378.6427 | -0.0009 | 0.04884 |
| | | | | | | | Blend | 2378.6415 | 0.0004 | |
| CO2-Xe-136 | 5 | 2 | 3 | 4 | 2 | 2 | 2378.6465 | 2378.6463 | 0.0002 | 0.02167 |
| CO2-Xe-136 | 7 | 0 | 7 | 6 | 0 | 6 | 2378.6465 | 2378.6464 | 0.0000 | 0.08075 |
| CO2-Xe-132 | 4 | 1 | 3 | 3 | 2 | 2 | 2378.6465 | 2378.6463 | 0.0002 | 0.02246 |
| | | | | | | | Blend | 2378.6464 | 0.0001 | |
| CO2-Xe-134 | 7 | 0 | 7 | 6 | 0 | 6 | 2378.6500 | 2378.6497 | 0.0003 | |
| CO2-Xe-131 | 7 | 0 | 7 | 6 | 0 | 6 | 2378.6536 | 2378.6546 | -0.0010 | 0.19322 |
| CO2-Xe-132 | 7 | 0 | 7 | 6 | 0 | 6 | 2378.6536 | 2378.6529 | 0.0006 | 0.24550 |
| | | | | | | | Blend | 2378.6536 | -0.0001 | |
| CO2-Xe-129 | 7 | 0 | 7 | 6 | 0 | 6 | 2378.6577 | 2378.6578 | -0.0001 | 0.23992 |
| CO2-Xe-129 | 7 | 2 | 6 | 6 | 2 | 5 | 2378.6577 | 2378.6574 | 0.0002 | 0.06689 |
| | | | | | | | Blend | 2378.6577 | -0.0001 | |
| CO2-Xe-129 | 8 | 2 | 6 | 7 | 2 | 5 | 2378.6742 | 2378.6744 | -0.0002 | 0.04450 |

| | | | | | | | | | |
|---|---|---|---|---|---|---|---|---|---|
| CO2-Xe-132 | 8 | 2 | 7 | 7 | 2 | 6 | 2378.6742 | 2378.6738 | 0.0005 | 0.04942 |
| | | | | | | | Blend | 2378.6740 | 0.0002 | |
| CO2-Xe-131 | 8 | 2 | 7 | 7 | 2 | 6 | 2378.6825 | 2378.6828 | -0.0003 | 0.04104 |
| CO2-Xe-129 | 4 | 1 | 4 | 3 | 2 | 1 | 2378.6825 | 2378.6835 | -0.0011 | 0.01389 |
| | | | | | | | Blend | 2378.6830 | -0.0005 | |
| CO2-Xe-136 | 8 | 0 | 8 | 7 | 0 | 7 | 2378.6944 | 2378.6944 | 0.0000 | 0.06767 |
| CO2-Xe-132 | 5 | 1 | 5 | 4 | 2 | 2 | 2378.6944 | 2378.6932 | 0.0012 | 0.03915 |
| | | | | | | | Blend | 2378.6939 | 0.0004 | |
| CO2-Xe-134 | 8 | 0 | 8 | 7 | 0 | 7 | 2378.6980 | 2378.6979 | 0.0000 | 0.07970 |
| CO2-Xe-129 | 8 | 2 | 7 | 7 | 2 | 6 | 2378.6980 | 2378.6983 | -0.0003 | 0.05584 |
| | | | | | | | Blend | 2378.6981 | -0.0001 | |
| CO2-Xe-132 | 8 | 0 | 8 | 7 | 0 | 7 | 2378.7023 | 2378.7015 | 0.0008 | 0.20528 |
| CO2-Xe-131 | 8 | 0 | 8 | 7 | 0 | 7 | 2378.7023 | 2378.7033 | -0.0010 | 0.16149 |
| | | | | | | | Blend | 2378.7023 | 0.0000 | |
| CO2-Xe-129 | 8 | 0 | 8 | 7 | 0 | 7 | 2378.7066 | 2378.7069 | -0.0003 | |
| CO2-Xe-132 | 9 | 2 | 8 | 8 | 2 | 7 | 2378.7117 | 2378.7117 | 0.0000 | |
| CO2-Xe-131 | 9 | 2 | 8 | 8 | 2 | 7 | 2378.7215 | 2378.7209 | 0.0006 | 0.03254 |
| CO2-Xe-132 | 5 | 1 | 4 | 4 | 2 | 3 | 2378.7215 | 2378.7224 | -0.0009 | 0.02889 |
| | | | | | | | Blend | 2378.7216 | -0.0001 | |
| CO2-Xe-134 | 9 | 0 | 9 | 8 | 0 | 8 | 2378.7439 | 2378.7436 | 0.0003 | |
| CO2-Xe-132 | 9 | 0 | 9 | 8 | 0 | 8 | 2378.7483 | 2378.7475 | 0.0008 | 0.16177 |
| CO2-Xe-131 | 9 | 0 | 9 | 8 | 0 | 8 | 2378.7483 | 2378.7495 | -0.0012 | 0.12719 |
| | | | | | | | Blend | 2378.7484 | -0.0001 | |
| CO2-Xe-129 | 9 | 0 | 9 | 8 | 0 | 8 | 2378.7531 | 2378.7534 | -0.0003 | |
| CO2-Xe-132 | 6 | 1 | 6 | 5 | 2 | 3 | 2378.7569 | 2378.7575 | -0.0006 | 0.04456 |
| CO2-Xe-131 | 10 | 2 | 9 | 9 | 2 | 8 | 2378.7569 | 2378.7572 | -0.0002 | 0.02425 |
| CO2-Xe-129 | 11 | 1 | 11 | 10 | 2 | 8 | 2378.7569 | 2378.7570 | -0.0001 | 0.01501 |
| | | | | | | | Blend | 2378.7573 | -0.0004 | |
| CO2-Xe-129 | 10 | 2 | 9 | 9 | 2 | 8 | 2378.7737 | 2378.7738 | -0.0001 | 0.03324 |
| CO2-Xe-131 | 6 | 1 | 6 | 5 | 2 | 3 | 2378.7737 | 2378.7727 | 0.0009 | 0.02975 |
| CO2-Xe-136 | 7 | 2 | 5 | 6 | 2 | 4 | 2378.7737 | 2378.7745 | -0.0009 | 0.02014 |
| | | | | | | | Blend | 2378.7736 | 0.0001 | |
| CO2-Xe-136 | 10 | 0 | 10 | 9 | 0 | 9 | 2378.7834 | 2378.7821 | 0.0013 | |
| CO2-Xe-132 | 10 | 0 | 10 | 9 | 0 | 9 | 2378.7917 | 2378.7908 | 0.0009 | 0.12033 |
| CO2-Xe-131 | 10 | 0 | 10 | 9 | 0 | 9 | 2378.7917 | 2378.7929 | -0.0012 | 0.09456 |
| | | | | | | | Blend | 2378.7917 | 0.0000 | |
| CO2-Xe-129 | 10 | 0 | 10 | 9 | 0 | 9 | 2378.7968 | 2378.7972 | -0.0005 | |

| | | | | | | | | | |
|---|---|---|---|---|---|---|---|---|---|
| CO2-Xe-129 | 11 | 0 | 11 | 10 | 0 | 10 | 2378.8381 | 2378.8383 | -0.0002 | |
| | | | | | | | | | | |
| CO2-Xe-132 | 12 | 0 | 12 | 11 | 0 | 11 | 2378.8710 | 2378.8689 | 0.0021 | 0.05667 |
| CO2-Xe-131 | 12 | 0 | 12 | 11 | 0 | 11 | 2378.8710 | 2378.8714 | -0.0005 | 0.04448 |
| CO2-Xe-132 | 7 | 1 | 6 | 6 | 2 | 5 | 2378.8710 | 2378.8712 | -0.0003 | 0.02897 |
| | | | | | | Blend | 2378.8703 | 0.0007 | |
| CO2-Xe-129 | 12 | 0 | 12 | 11 | 0 | 11 | 2378.8759 | 2378.8766 | -0.0007 | 0.05491 |
| CO2-Xe-132 | 8 | 2 | 6 | 7 | 2 | 5 | 2378.8759 | 2378.8771 | -0.0013 | 0.03943 |
| | | | | | | Blend | 2378.8768 | -0.0009 | |

Table A-5. Observed and fitted $(01^11) \leftarrow (01^10)$ hot band transitions of $CO_2$-Xe (cm$^{-1}$).

Int = calculated relative intensity, used to weight the blended lines.

Transitions with $Ka'' = 0$ and 2 belong to the in-plane band.

Transitions with $Ka'' = 1$ and 3 belong to the out-of-plane band.

| $J'$ | $Ka'$ | $Kc'$ | $J''$ | $Ka''$ | $Kc''$ | Observed | Calculated | Obs - Calc | Int |
|---|---|---|---|---|---|---|---|---|---|
| 3 | 2 | 1 | 4 | 3 | 2 | 2333.1193 | 2333.1195 | -0.0002 | 0.00010 |
| 3 | 2 | 2 | 4 | 3 | 1 | 2333.1193 | 2333.1192 | 0.0000 | 0.00010 |
| | | | | | | Blend | 2333.1193 | -0.0001 | |
| 2 | 2 | 0 | 3 | 3 | 1 | 2333.1453 | 2333.1453 | 0.0000 | 0.00015 |
| 2 | 2 | 1 | 3 | 3 | 0 | 2333.1453 | 2333.1452 | 0.0001 | 0.00015 |
| | | | | | | Blend | 2333.1453 | 0.0001 | |
| 6 | 1 | 5 | 7 | 2 | 6 | 2333.6121 | 2333.6121 | 0.0000 | |
| 5 | 1 | 5 | 6 | 2 | 4 | 2333.6641 | 2333.6638 | 0.0003 | |
| 5 | 1 | 4 | 6 | 2 | 5 | 2333.6712 | 2333.6712 | 0.0000 | |
| 4 | 1 | 4 | 5 | 2 | 3 | 2333.7270 | 2333.7269 | 0.0001 | |
| 4 | 1 | 3 | 5 | 2 | 4 | 2333.7315 | 2333.7313 | 0.0003 | |
| 3 | 1 | 3 | 4 | 2 | 2 | 2333.7910 | 2333.7899 | 0.0010 | 0.00179 |
| 3 | 1 | 2 | 4 | 2 | 3 | 2333.7910 | 2333.7922 | -0.0012 | 0.00180 |
| | | | | | | Blend | 2333.7911 | -0.0001 | |
| 2 | 1 | 1 | 3 | 2 | 2 | 2333.8535 | 2333.8540 | -0.0005 | 0.00187 |
| 2 | 1 | 2 | 3 | 2 | 1 | 2333.8535 | 2333.8531 | 0.0005 | 0.00186 |
| | | | | | | Blend | 2333.8535 | 0.0000 | |
| 1 | 1 | 1 | 2 | 2 | 0 | 2333.9165 | 2333.9165 | 0.0000 | 0.00193 |
| 1 | 1 | 0 | 2 | 2 | 1 | 2333.9165 | 2333.9168 | -0.0003 | 0.00193 |
| | | | | | | Blend | 2333.9166 | -0.0002 | |
| 4 | 1 | 3 | 4 | 2 | 2 | 2334.0548 | 2334.0546 | 0.0003 | 0.00175 |
| 5 | 1 | 5 | 5 | 2 | 4 | 2334.0548 | 2334.0557 | -0.0009 | 0.00171 |
| | | | | | | Blend | 2334.0551 | -0.0003 | |
| 5 | 1 | 4 | 5 | 2 | 3 | 2334.0589 | 2334.0588 | 0.0001 | 0.00176 |
| 6 | 1 | 6 | 6 | 2 | 5 | 2334.0589 | 2334.0593 | -0.0004 | 0.00155 |
| | | | | | | Blend | 2334.0591 | -0.0001 | |
| 6 | 1 | 5 | 6 | 2 | 4 | 2334.0634 | 2334.0636 | -0.0002 | 0.00160 |
| 7 | 1 | 7 | 7 | 2 | 6 | 2334.0634 | 2334.0632 | 0.0002 | 0.00130 |
| | | | | | | Blend | 2334.0634 | -0.0001 | |
| 7 | 1 | 6 | 7 | 2 | 5 | 2334.0683 | 2334.0688 | -0.0005 | 0.00136 |
| 8 | 1 | 8 | 8 | 2 | 7 | 2334.0683 | 2334.0672 | 0.0011 | 0.00102 |
| | | | | | | Blend | 2334.0681 | 0.0002 | |
| 8 | 1 | 7 | 8 | 2 | 6 | 2334.0747 | 2334.0744 | 0.0003 | 0.00107 |
| 10 | 1 | 10 | 10 | 2 | 9 | 2334.0747 | 2334.0749 | -0.0002 | 0.00051 |
| | | | | | | Blend | 2334.0746 | 0.0002 | |
| 9 | 1 | 8 | 9 | 2 | 7 | 2334.0803 | 2334.0803 | 0.0000 | |
| 10 | 1 | 9 | 10 | 2 | 8 | 2334.0866 | 2334.0865 | 0.0001 | |
| 7 | 0 | 7 | 8 | 1 | 8 | 2334.3139 | 2334.3145 | -0.0005 | |

| | | | | | | | | |
|---|---|---|---|---|---|---|---|---|
| 6 | 0 | 6 | 7 | 1 | 7 | 2334.3712 | 2334.3710 | 0.0003 | |
| 5 | 1 | 5 | 4 | 2 | 2 | 2334.3796 | 2334.3790 | 0.0006 | |
| 5 | 1 | 4 | 4 | 2 | 3 | 2334.3843 | 2334.3843 | 0.0000 | |
| 5 | 0 | 5 | 6 | 1 | 6 | 2334.4288 | 2334.4287 | 0.0001 | |
| 6 | 1 | 6 | 5 | 2 | 3 | 2334.4466 | 2334.4469 | -0.0003 | |
| 6 | 1 | 5 | 5 | 2 | 4 | 2334.4550 | 2334.4555 | -0.0005 | |
| 4 | 0 | 4 | 5 | 1 | 5 | 2334.4881 | 2334.4879 | 0.0002 | |
| 3 | 0 | 3 | 4 | 1 | 4 | 2334.5491 | 2334.5488 | 0.0003 | |
| 2 | 0 | 2 | 3 | 1 | 3 | 2334.6119 | 2334.6116 | 0.0004 | |
| 10 | 0 | 10 | 10 | 1 | 9 | 2334.6373 | 2334.6369 | 0.0004 | |
| 9 | 0 | 9 | 9 | 1 | 8 | 2334.6693 | 2334.6696 | -0.0003 | |
| 1 | 0 | 1 | 2 | 1 | 2 | 2334.6760 | 2334.6762 | -0.0002 | |
| 8 | 0 | 8 | 8 | 1 | 7 | 2334.6982 | 2334.6984 | -0.0002 | |
| 7 | 0 | 7 | 7 | 1 | 6 | 2334.7236 | 2334.7236 | 0.0000 | |
| 5 | 0 | 5 | 5 | 1 | 4 | 2334.7644 | 2334.7642 | 0.0002 | |
| 4 | 0 | 4 | 4 | 1 | 3 | 2334.7796 | 2334.7798 | -0.0002 | |
| 3 | 0 | 3 | 3 | 1 | 2 | 2334.7922 | 2334.7924 | -0.0002 | |
| 2 | 0 | 2 | 2 | 1 | 1 | 2334.8016 | 2334.8019 | -0.0003 | |
| 1 | 0 | 1 | 1 | 1 | 0 | 2334.8079 | 2334.8082 | -0.0004 | |
| 8 | 1 | 8 | 9 | 0 | 9 | 2334.8955 | 2334.8954 | 0.0000 | |
| 7 | 1 | 7 | 8 | 0 | 8 | 2334.9644 | 2334.9646 | -0.0002 | |
| 6 | 1 | 6 | 7 | 0 | 7 | 2335.0325 | 2335.0331 | -0.0006 | |
| 5 | 1 | 5 | 6 | 0 | 6 | 2335.1012 | 2335.1011 | 0.0001 | |
| 4 | 1 | 4 | 5 | 0 | 5 | 2335.1685 | 2335.1686 | 0.0000 | |
| 5 | 0 | 5 | 4 | 1 | 4 | 2335.1813 | 2335.1805 | 0.0007 | |
| 3 | 1 | 3 | 4 | 0 | 4 | 2335.2359 | 2335.2356 | 0.0003 | |
| 6 | 0 | 6 | 5 | 1 | 5 | 2335.2585 | 2335.2585 | 0.0000 | |
| 2 | 1 | 2 | 3 | 0 | 3 | 2335.3024 | 2335.3022 | 0.0002 | |
| 7 | 0 | 7 | 6 | 1 | 6 | 2335.3370 | 2335.3372 | -0.0002 | |
| 1 | 1 | 1 | 2 | 0 | 2 | 2335.3687 | 2335.3685 | 0.0001 | |
| | | | | | | | | | |
| 4 | 1 | 3 | 4 | 0 | 4 | 2335.5005 | 2335.5002 | 0.0002 | 0.01084 |
| 5 | 1 | 4 | 5 | 0 | 5 | 2335.5005 | 2335.5004 | 0.0000 | 0.01042 |
| 3 | 1 | 2 | 3 | 0 | 3 | 2335.5005 | 2335.5002 | 0.0003 | 0.01021 |
| 6 | 1 | 5 | 6 | 0 | 6 | 2335.5005 | 2335.5009 | -0.0004 | 0.00923 |
| 2 | 1 | 1 | 2 | 0 | 2 | 2335.5005 | 2335.5002 | 0.0003 | 0.00842 |
| 1 | 1 | 0 | 1 | 0 | 1 | 2335.5005 | 2335.5002 | 0.0002 | 0.00556 |
| | | | | | | Blend | 2335.5004 | 0.0001 | |
| 11 | 1 | 10 | 11 | 0 | 11 | 2335.5096 | 2335.5095 | 0.0000 | |
| 1 | 1 | 1 | 0 | 0 | 0 | 2335.5657 | 2335.5657 | 0.0000 | |
| 2 | 1 | 2 | 1 | 0 | 1 | 2335.6306 | 2335.6309 | -0.0003 | |
| 3 | 1 | 3 | 2 | 0 | 2 | 2335.6957 | 2335.6958 | -0.0001 | |
| 4 | 1 | 4 | 3 | 0 | 3 | 2335.7605 | 2335.7604 | 0.0001 | |
| 5 | 1 | 5 | 4 | 0 | 4 | 2335.8248 | 2335.8247 | 0.0001 | |
| 6 | 1 | 6 | 5 | 0 | 5 | 2335.8887 | 2335.8885 | 0.0002 | |
| 7 | 1 | 7 | 6 | 0 | 6 | 2335.9520 | 2335.9519 | 0.0000 | |
| 8 | 1 | 8 | 7 | 0 | 7 | 2336.0149 | 2336.0148 | 0.0001 | |
| 9 | 1 | 9 | 8 | 0 | 8 | 2336.0770 | 2336.0771 | -0.0001 | 0.00360 |
| 9 | 1 | 9 | 8 | 0 | 8 | 2336.0770 | 2336.0771 | -0.0001 | 0.00360 |
| | | | Blend | | | | 2336.0771 | -0.0001 | |
| 9 | 2 | 7 | 9 | 1 | 8 | 2336.2230 | 2336.2232 | -0.0002 | |
| | | | | | | | | | |
| 8 | 2 | 6 | 8 | 1 | 7 | 2336.2308 | 2336.2298 | 0.0010 | 0.00044 |
| 2 | 2 | 0 | 2 | 1 | 1 | 2336.2308 | 2336.2311 | -0.0003 | 0.00046 |
| | | | Blend | | | | 2336.2305 | 0.0003 | |
| 7 | 2 | 5 | 7 | 1 | 6 | 2336.2354 | 2336.2348 | 0.0006 | 0.00056 |
| 3 | 2 | 1 | 3 | 1 | 2 | 2336.2354 | 2336.2349 | 0.0005 | 0.00068 |
| | | | Blend | | | | 2336.2349 | 0.0005 | |
| 5 | 2 | 3 | 5 | 1 | 4 | 2336.2382 | 2336.2388 | -0.0006 | 0.00076 |
| 6 | 2 | 4 | 6 | 1 | 5 | 2336.2382 | 2336.2379 | 0.0003 | 0.00068 |
| 4 | 2 | 2 | 4 | 1 | 3 | 2336.2382 | 2336.2377 | 0.0004 | 0.00077 |
| | | | Blend | | | | 2336.2382 | 0.0000 | |
| 2 | 2 | 1 | 2 | 1 | 2 | 2336.2462 | 2336.2463 | -0.0001 | |
| 3 | 2 | 2 | 3 | 1 | 3 | 2336.2646 | 2336.2649 | -0.0004 | |
| 4 | 2 | 3 | 4 | 1 | 4 | 2336.2869 | 2336.2871 | -0.0002 | |
| 5 | 2 | 4 | 5 | 1 | 5 | 2336.3116 | 2336.3117 | -0.0001 | |
| 6 | 2 | 5 | 6 | 1 | 6 | 2336.3381 | 2336.3382 | -0.0001 | |
| 2 | 2 | 1 | 1 | 1 | 0 | 2336.3785 | 2336.3783 | 0.0002 | |
| 2 | 2 | 0 | 1 | 1 | 1 | 2336.3835 | 2336.3834 | 0.0000 | |
| 8 | 2 | 7 | 8 | 1 | 8 | 2336.3951 | 2336.3949 | 0.0001 | |
| 3 | 2 | 2 | 2 | 1 | 1 | 2336.4554 | 2336.4553 | 0.0002 | |
| 3 | 2 | 1 | 2 | 1 | 2 | 2336.4710 | 2336.4707 | 0.0003 | |
| 4 | 2 | 3 | 3 | 1 | 2 | 2336.5300 | 2336.5307 | -0.0007 | |